\documentclass[twocolumn]{revtex4}

\usepackage{amsmath, amssymb}
\usepackage{natbib}
\usepackage[colorlinks, 
linkcolor=blue,
urlcolor=blue,
citecolor=blue,
bookmarksopen=false]{hyperref}

\usepackage{amsmath}
\usepackage{graphicx, xcolor}
\usepackage{caption}
\usepackage{amsmath}
\usepackage{float}

\def\iotabar{\lower3pt\hbox{$\mathchar'26$}\mkern-7mu\iota}
\newcommand {\aplt} {\ {\raise-.5ex\hbox{$\buildrel<\over\sim$}}\ }

\newcommand{\dotcross}{ \raise 0.65ex\hbox{${\scriptstyle {{_{\displaystyle \cdot}}\atop\times}}$} }
\newcommand{\crossdot}{ \raise 0.5ex\hbox{${\scriptstyle {{_\times}\atop{\displaystyle \cdot}}}$} }

\newcommand{\sumsig}{ \raise -1.3ex\hbox{${{\displaystyle \sum}\atop{\scriptstyle \sigma}}$} }
\newcounter{appnumb}

\begin{document}

\title{Combination of quasi-isodynamic and piecewise omnigenous magnetic fields}

\author{J.~L. Velasco$^1$\email{joseluis.velasco@ciemat.es}, I. Calvo$^1$, V. Fern\'andez-Pacheco$^1$, H. Liu$^2$,  M. Padidar$^3$, E. S\'anchez$^1$, G. Yu$^2$, C. Zhu$^2$}

\keywords{stellarator design, piecewise omnigeneity, quasi-isodynamicity, neoclassical transport}

\affiliation{$^1$ Laboratorio Nacional de Fusi\'on, CIEMAT, 28040, Madrid, Spain\\
$^2$ University of Science and Technology of China, Hefei, Anhui 230026, China\\
$^3$ Center for Computational Mathematics, Flatiron Institute, New York, NY 10010, USA}


\date{\today}

\begin{abstract}
  
Due to their simultaneous optimization for radial and parallel neoclassical transport, quasi-isodynamic fields have been the main choice of stellarator magnetic configuration for most fusion reactor candidates in recent years.  However, achieving a high degree of quasi-isodynamicity often comes at the cost of a strong shaping of the flux surfaces of the stellarator and complex coil geometries. In this work, the concepts of quasi-isodynamicity and piecewise omnigenity are combined to form QI-pwO fields. These fields are quasi-isodynamic in the low-field region of the magnetic surface, whereas they significantly depart from quasi-isodynamicity in the high-field region without sacrificing the neoclassical transport properties of quasi-isodynamic fields. This departure could make it easier to integrate the optimization of neoclassical transport with other physical and technological aspects of a stellarator reactor.
    
\end{abstract}

\maketitle

\section{Introduction}

Recent theoretical and numerical advances in magnetic confinement fusion have led to the design of a number of approximately omnigenous stellarator magnetic configurations, see e.g. \cite{henneberg2019qa,plunk2019direct,kinoshita2019cfqs,bader2020wistell,landreman2022preciseQS,landreman2022mapping,camachomata2022direct,jorge2022qi,sanchez2023qi,goodman2023qi,dudt2023omni,bindel2023direct,goodman2024squids,liu2025omni,lion2025stellaris,hegna2025infinity2}, some of which have been proposed as the basis of stellarator reactors~\cite{sanchez2023qi,goodman2024squids,lion2025stellaris,hegna2025infinity2,swanson2026helios}. Omnigenity~\cite{hall1975omni} is a property of toroidal magnetic fields (specifically, of the nested \textit{flux surfaces} that form the magnetic configuration) that ensures that the neoclassical transport of the stellarator, associated to the combination of radial drifts and collisions \cite{helander2002collisional}, is relatively benign, as in an axisymmetric tokamak. Otherwise, in a generic stellarator, neoclassical transport can severely limit the achievable plasma pressure.

\begin{figure}
\includegraphics[angle=0,width=.49\columnwidth]{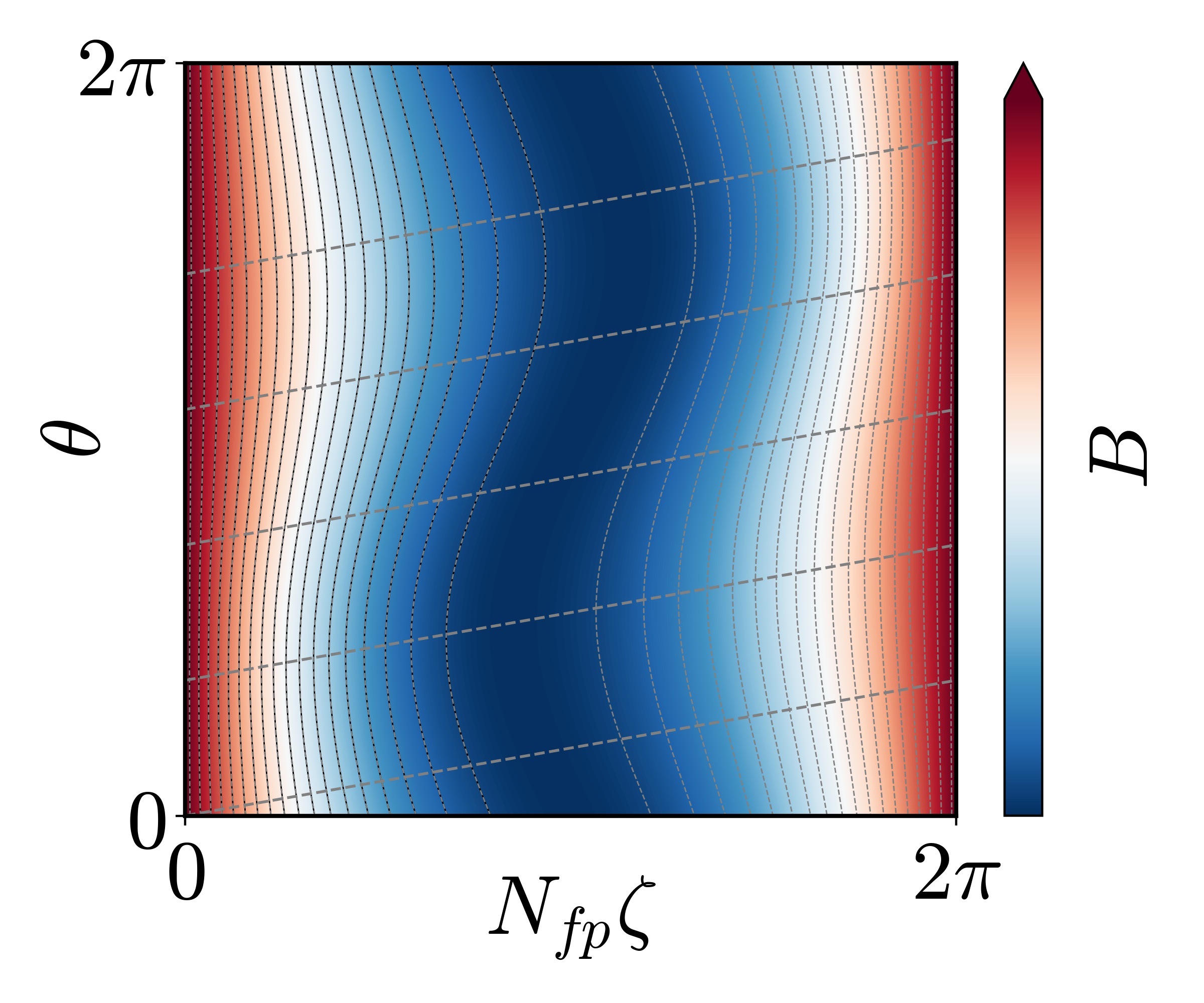}
\includegraphics[angle=0,width=.49\columnwidth]{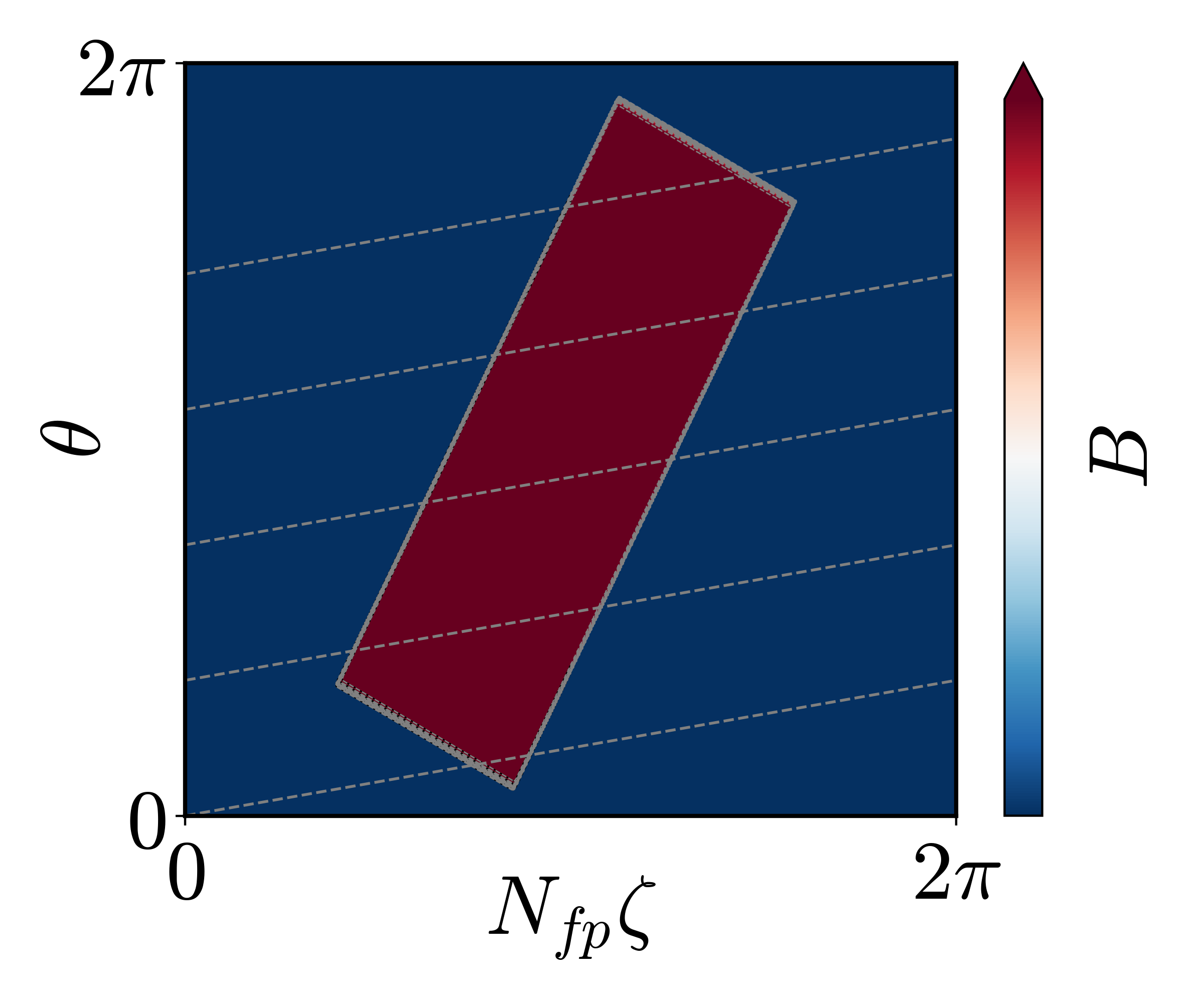}
\caption{Magnetic field strength of a QI field (left) and of a pwO field far from quasi-isodynamicity (right).\label{FIG_EXAMPLES}}
\end{figure}

Omnigenity requires a careful tailoring of the variation of the magnetic field strength $B$ on a flux surface \cite{cary1997omni,parra2015omni,landreman2012omni}. First, all the contours of constant $B$ must close toroidally, poloidally or helically. Additionally, the distance between two contours corresponding to the same value of $B$, measured along the field lines, must be independent of the field line. Interestingly, in most of the reactor candidates~\cite{sanchez2023qi,goodman2024squids,lion2025stellaris,hegna2025infinity2}, the magnetic field has been chosen to be approximately quasi-isodynamic (QI), a subclass of omnigenous fields with $B$-contours that close poloidally. QI fields are particularly attractive since they guarantee zero \textit{bootstrap} current at low collisionality~\cite{helander2009bootstrap}, which enables compatibility of the magnetic configuration with the use of an island divertor, see e.g.~\cite{sunnpedersen2019divertor}, for energy and particle exhaust. An example of a QI field is shown in figure~\ref{FIG_EXAMPLES} (left).

Exact quasi-isodynamic fields are not analytic \cite{cary1997omni}, and can therefore not be achieved even on a single flux surface of a stellarator. For this reason, constructions of QI equilibria usually break quasi-isodynamicity in the region of maximum $B$, $B_{max}$ \cite{plunk2019direct,rodriguez2023qi2}. Even so, the required topology for the $B$-contours often leads to complicated shapes of the flux surfaces, particularly with high elongation~\cite{camachomata2022direct}; this may, in turn, result in complicated coils. The surface elongation effect may be even more pronounced if the \textit{radial} variation of $B$ (i.e., the variation across flux surfaces) is additionally tailored to minimize fast ion losses and some types of turbulent transport, see e.g.~\cite{rodriguez2024maxj}.

Piecewise omnigenous (pwO) fields~\cite{velasco2024pwO} constitute a new, broader, family of optimized stellarator fields displaying levels of neoclassical confinement similar to that of tokamaks and omnigenous stellarators. They are not subject to some of the constraints that must be fulfilled by omnigenous fields~\cite{cary1997omni,parra2015omni}. For instance, the contours of constant magnetic field strength $B$ on do not need to close toroidally, poloidally or helically.  This means that they do not display a preferred \textit{helicity}. By relaxing the constraints on omnigenous fields, pwO fields have the potential to radically broaden the space of accessible reactor-relevant magnetic configurations. For the simplest or \textit{prototypical} case of pwO field, $B=B_\mathrm{max}$ inside a parallelogram-shaped region of the flux surface, and $B=B_\mathrm{min}$ outside it. An example of such pwO field is shown in figure~\ref{FIG_EXAMPLES} (right).

Recently, several works have pushed the concept of pwO fields, showing that they can achieve similar confinement properties to QI fields and are a viable alternative. First, in~\cite{calvo2025pwO}, it has been shown that prototypical pwO fields can be designed to have zero bootstrap current at low collisionality, thus being simultaneously optimized for reduced radial and parallel neoclassical transport, a set of characteristics that was previously thought to be specific of QI fields. In addition, by means of numerical optimization, it has been demonstrated that such pwO fields can be the basis of a magnetohydrodynamic (MHD) equilibrium \cite{fernandezpacheco2026pw1} that fulfils the standard set of physics criteria usually required for a stellarator reactor \cite{sanchez2023qi,goodman2024squids,lion2025stellaris,hegna2025infinity2}. In parallel, in~\cite{velasco2025parapwO}, the configuration space of pwO fields has been explored in detail, beyond the prototypical fields of~\cite{velasco2024pwO,calvo2025pwO}. In particular, pwO fields have been proposed that are approximately omnigenous, in the sense that outside the parallelogram-shaped region of  $B_\mathrm{max}$, the contours of constant $B$ close toroidally, poloidally or helically and, with the appropriate distance between them~\cite{cary1997omni,landreman2012omni}, do not to give rise to a large neoclassical transport.  Finally, a novel optimization method \cite{liu2025omni} has yielded MHD equilibria with approximately pwO magnetic fields whose $B_\mathrm{min}$-contours close poloidally.

Leveraging these works, we would like to obtain magnetic fields that depart significantly from quasi-isodynamicity in the high-field region of the flux surface and still present, thanks to piecewise omnigenity, tokamak-like levels of radial neoclassical transport and, additionally, negligible bootstrap current. The focus of this work is the characterization of the appropriate pwO deviations from quasi-isodynamicity. The expectation is that departing from quasi-isodynamicity, and thus relaxing the constraints imposed by it, could lead to less elongated shapes and/or simpler coils. Some preliminary indications of this are presented in \cite{liu2025omni}.

In order to fulfil our goal, we will design $B$ such that deeply trapped particles ($\mathcal{E}/\mu\gtrsim B_\mathrm{min}$, with $\mathcal{E}$ and $\mu$ the particle energy and magnetic moment, respectively, and $B_\mathrm{min}$ the minimum of $B$) will behave as in a QI field, bouncing back and forth between bounce points that lie on $B$-contours that close poloidally as in figure~\ref{FIG_EXAMPLES} (left). Conversely, barely trapped particles ($\mathcal{E}/\mu \lesssim B_\mathrm{max}$) will behave as in a pwO field like those of~\cite{calvo2025pwO} or figure~\ref{FIG_EXAMPLES} (right), bouncing back and forth between bounce points that lie on the sides of a parallelogram. We will illustrate this with an example in section~\ref{SEC_MODEL}, see figure~\ref{FIG_MODEL}. With some abuse of terminology (since, strictly speaking, quasi-isodynamicity is a property of the flux surface as a whole) one can refer to the pwO magnetic field described in the previous paragraph as a \textit{combination} of pwO and QI pieces. We will thus speak of QI-pwO fields. In fact, QI fields can be smoothly deformed into QI-pwO fields, and vice versa, as we will illustrate in section~\ref{SEC_SCANS} (see also~\cite{escoto2025qi}). 


Finally, the concept of QI-pwO field can clarify and systematize the notion of \textit{quasi-omnigenity}, see e.g.~\cite{spong1998jstar}. The term quasi-omnigenous has been loosely employed (see~\cite{mynick2006opt}, where it is applied to both Wendelstein 7-X and LHD) to describe magnetic configurations that are only approximately omnigenous. In many cases, it refers to configurations with good transport properties even though the conditions of omnigenity (or quasi-isodynamicity) are fulfilled in the region of the flux surface with $B \gtrsim B_\mathrm{min}$ but not where $B\lesssim B_\mathrm{max}$. Aside from its lack of rigor, this definition also presents practical difficulties, since there exists a wide variety of approximately omnigenous fields, many of which are not compatible with a reactor scenario. In this work, we will characterize the deviations from quasi-isodynamicity at $B\lesssim B_\mathrm{max}$ that cause negligible radial neoclassical transport and bootstrap current. By doing so, we will identify the most reactor-relevant subset of quasi-omnigenous fields.

The rest of this paper is organized as follows. Section~\ref{SEC_MODEL} presents a model for a QI-pwO field and compares it to W7-X. Then, in section \ref{SEC_SCANS}, we perform scans in the parameters of the model. Section~\ref{SEC_CONFS} shows MHD equilibria close to being QI-pwO, at least in some flux surfaces. The discussion comes in section~\ref{SEC_DISC}.

\section{A model for a QI-pwO field}\label{SEC_MODEL}

We now present a model for ideal QI-pwO fields. For periodicity $N_{fp}$ and rotational transform $\iota$, the variation of the magnetic field strength as a function of the toroidal and poloidal Boozer angles, $\zeta$ and $\theta$, is
\begin{eqnarray}
B(\theta,\zeta) &=&B_{0}\, f_{pwO}(\theta,\zeta)\, f_{QI}(\theta,\zeta)\,,\label{EQ_QIPWO}\\
f_{pwO}(\theta,\zeta)&=&1+\epsilon\, e^{-\left(\frac{\zeta-\zeta_c+t_1(\theta-\theta_c)}{w_1}\right)^{2p}-\left(\frac{\theta-\theta_c+t_2(\zeta-\zeta_c)}{w_2}\right)^{2p}}\,,\label{EQ_PWO}\\ 
w_1&=&\frac{\pi}{N_{fp}} \frac{1-t_1t_2}{1+t_2/\iota}\,,\label{EQ_W1}\\
w_2 &=&\pi\,,\label{EQ_W2}\\
f_{QI}(\theta,\zeta)&=&
\begin{cases}
1~~~~~\mathrm{if}~\zeta-\zeta_\mathrm{min}(\theta+\iota(\zeta_\mathrm{min}-\zeta))>\Delta\zeta\,, \\
~~~~~~\mathrm{or}~\zeta_\mathrm{min}(\theta+\iota(\zeta_\mathrm{min}-\zeta)
)-\zeta>\Delta\zeta\,,\\  
1-0.5\epsilon\left[1+\cos\left(\frac{\pi(\zeta_\mathrm{min}(\theta+\iota(\zeta_\mathrm{min}-\zeta))-\zeta)}{\Delta\zeta}\right)\right]\\~\mathrm{elsewhere}\,,
\end{cases}\label{EQ_QI}\\
\zeta_\mathrm{min}(\theta)&=&t_1\sin(\theta)\,,\label{EQ_ZMIN}\\
t_2 &=&-\frac{1}{t_1}\,.\label{EQ_T2}
\end{eqnarray}

Here, equation (\ref{EQ_PWO}) parametrizes the (approximately, for finite $p$) pwO region of the magnetic field (quantities $B_0$, $\epsilon$, $w_i$, $t_i$, $\zeta_c$ and $\theta_c$ parametrize $B$-contours that are approximately parallelograms~\cite{velasco2025parapwO}). Equation (\ref{EQ_W1}) sets $w_1$ as a function of the rest of parameters to guarantee tokamak-like radial transport in the limit ${p\to\infty}$~\cite{velasco2024pwO}. In the same limit, the value of $w_2$ given by equation (\ref{EQ_W2}) guarantees zero bootstrap current at low collisionality~\cite{calvo2025pwO}. Equation (\ref{EQ_QI}) describes a QI field constrained to a region of the flux surface at a distance (along the field line) $\pm\Delta\zeta$ from the $B_\mathrm{min}$-contour, $\zeta_\mathrm{min}(\theta)$, given by equation (\ref{EQ_ZMIN}). The poloidal dependence of  $\zeta_\mathrm{min}$, combined with equation (\ref{EQ_T2}), provides a reasonable alignment of the sides of the characteristic parallelogram of the pwO piece with the $B$-contours of the QI region of the field. Finally, in our model, the bounce points lie on poloidally-closed $B$-contours for $B_0(1-\epsilon)\leq\mathcal{E}/\mu\leq B_0$, and on a parallelogram-shaped $B$-contour for trapped particles with $B_0<\mathcal{E}/\mu\leq B_0(1+\epsilon)$. The relative size, in velocity space, of these two regions could be easily changed by modifying equations (\ref{EQ_PWO}) and (\ref{EQ_QI}).

Altogether, in the limit ${p\to\infty}$ (and for sufficiently small $\Delta\zeta$), equations~(\ref{EQ_QIPWO}), (\ref{EQ_PWO}), (\ref{EQ_W1}), (\ref{EQ_W2}). (\ref{EQ_QI}), (\ref{EQ_ZMIN}), and (\ref{EQ_T2}) represent a QI-pwO field as defined in the introduction. As such, although not all its $B$-contours close poloidally, it is expected to have the same level of radial and parallel neoclassical transport than a QI field. For a finite value of $p$ and $\Delta\zeta \approx 2\pi/N_{fp}-w_1$, the field resembles the pwO case of \cite{liu2025omni}, and is expected to show levels of neoclassical transport that are compatible with a reactor.

\begin{figure}
\includegraphics[angle=0,width=.49\columnwidth]{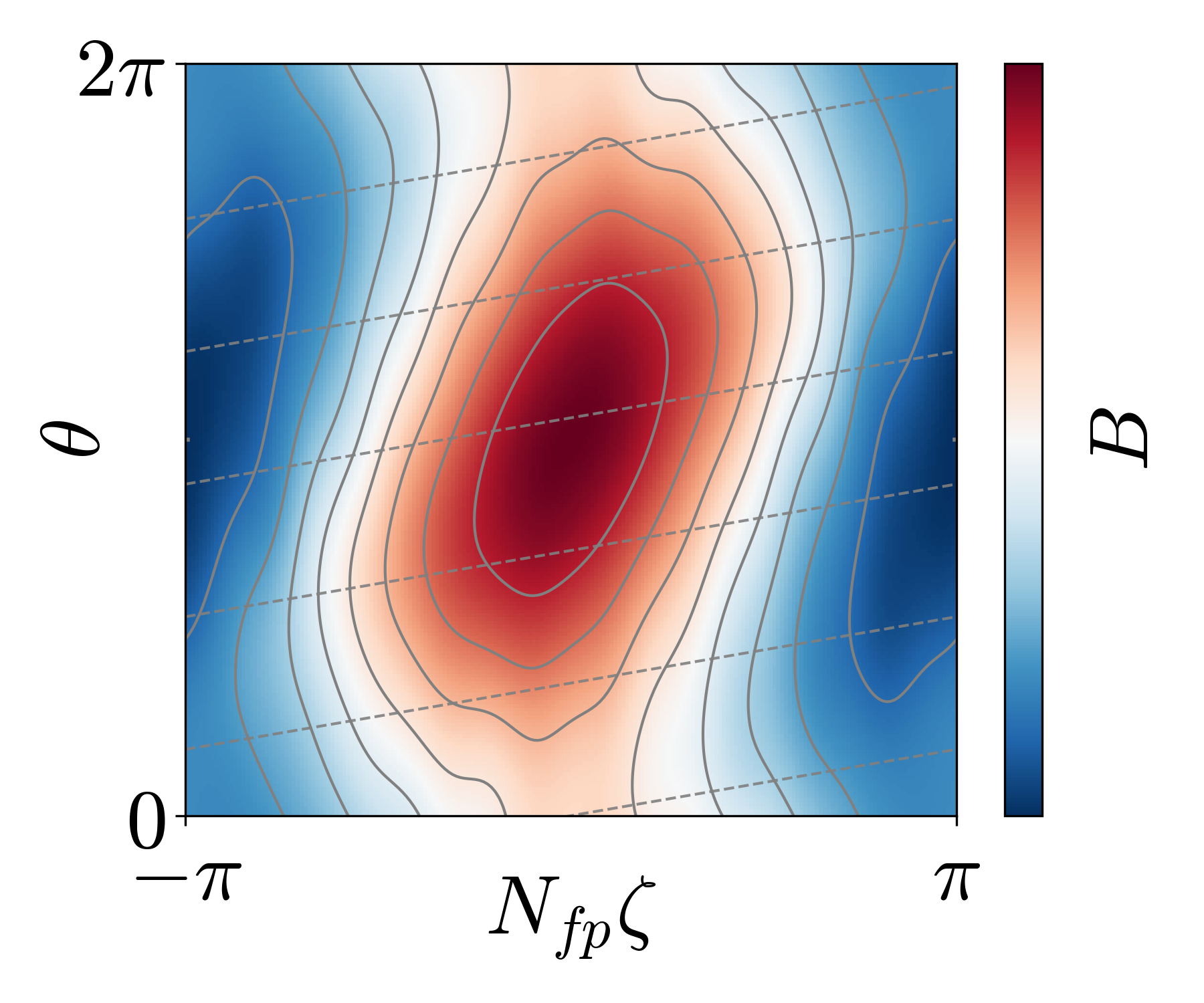}
\includegraphics[angle=0,width=.49\columnwidth]{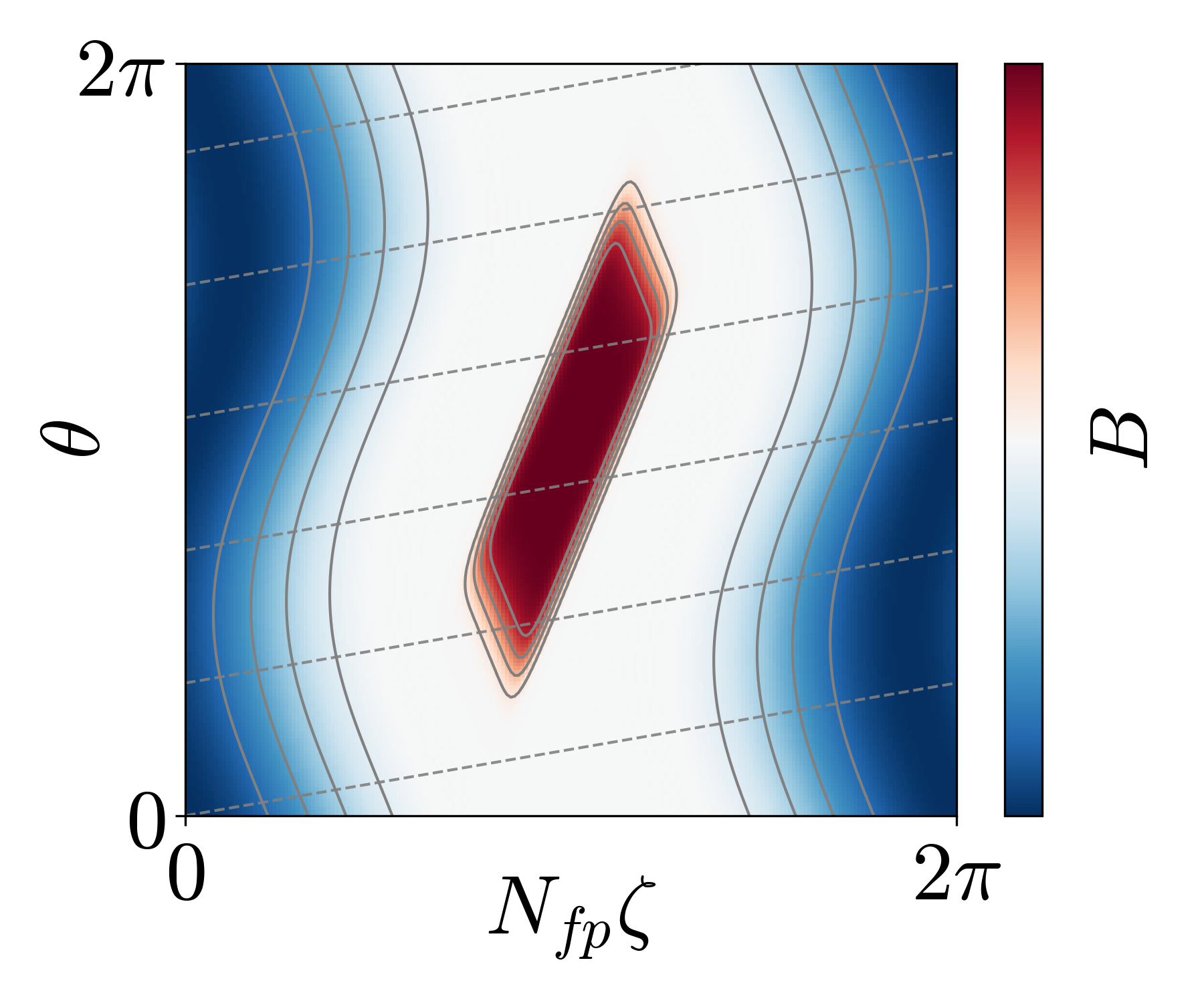}
\caption{Magnetic field strength of W7-X high-mirror at $s=0.25$ (left) and a QI-pwO field (right). For W7-X, the toroidal range is shifted by half a period, since $B=B_\mathrm{max}$ lies at $\zeta=0$ instead of $\zeta=\pi/N_{fp}$.\label{FIG_MODEL}}
\end{figure}

\begin{figure}
\includegraphics[angle=0,width=.49\columnwidth]{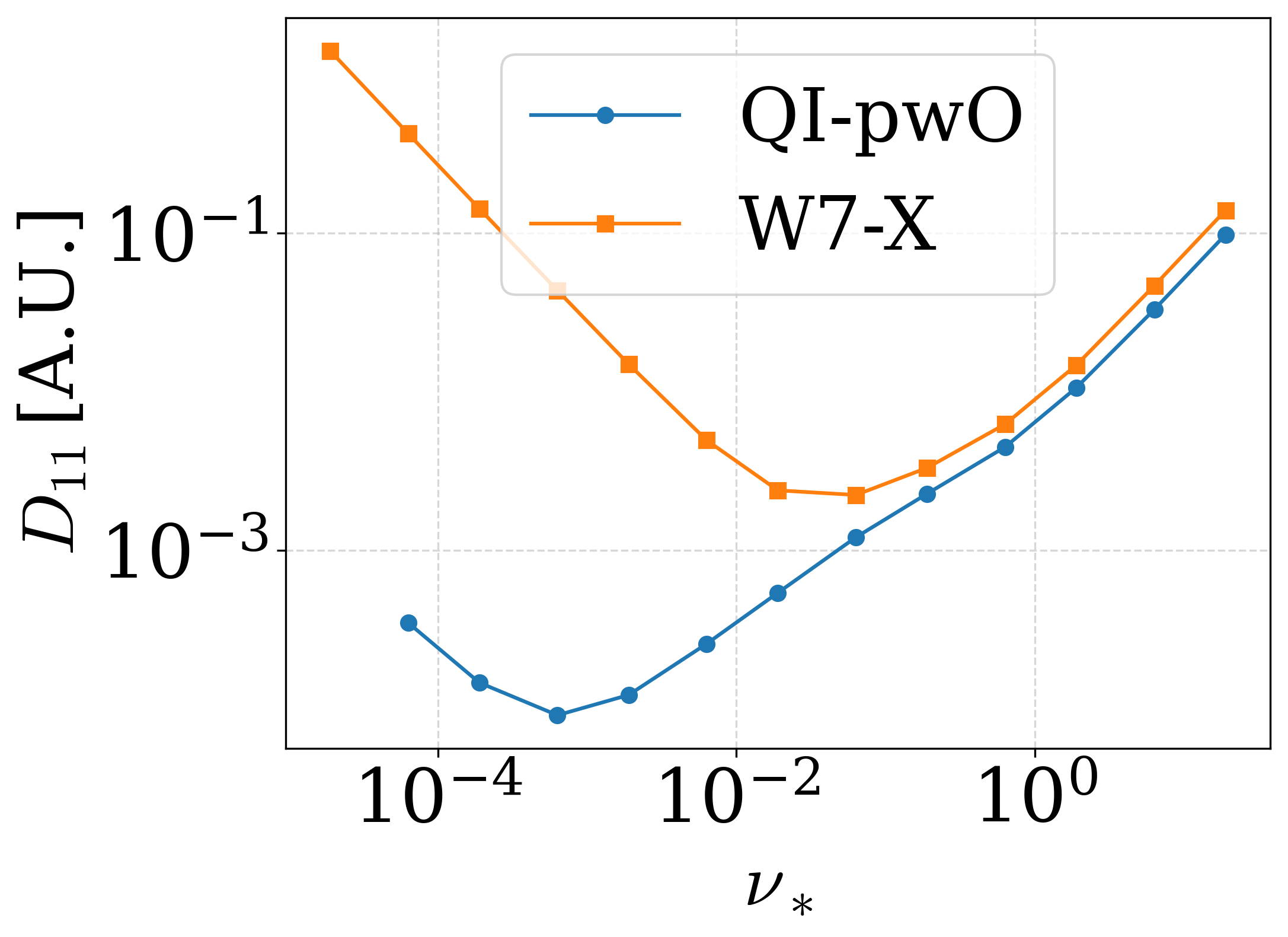}  
\includegraphics[angle=0,width=.49\columnwidth]{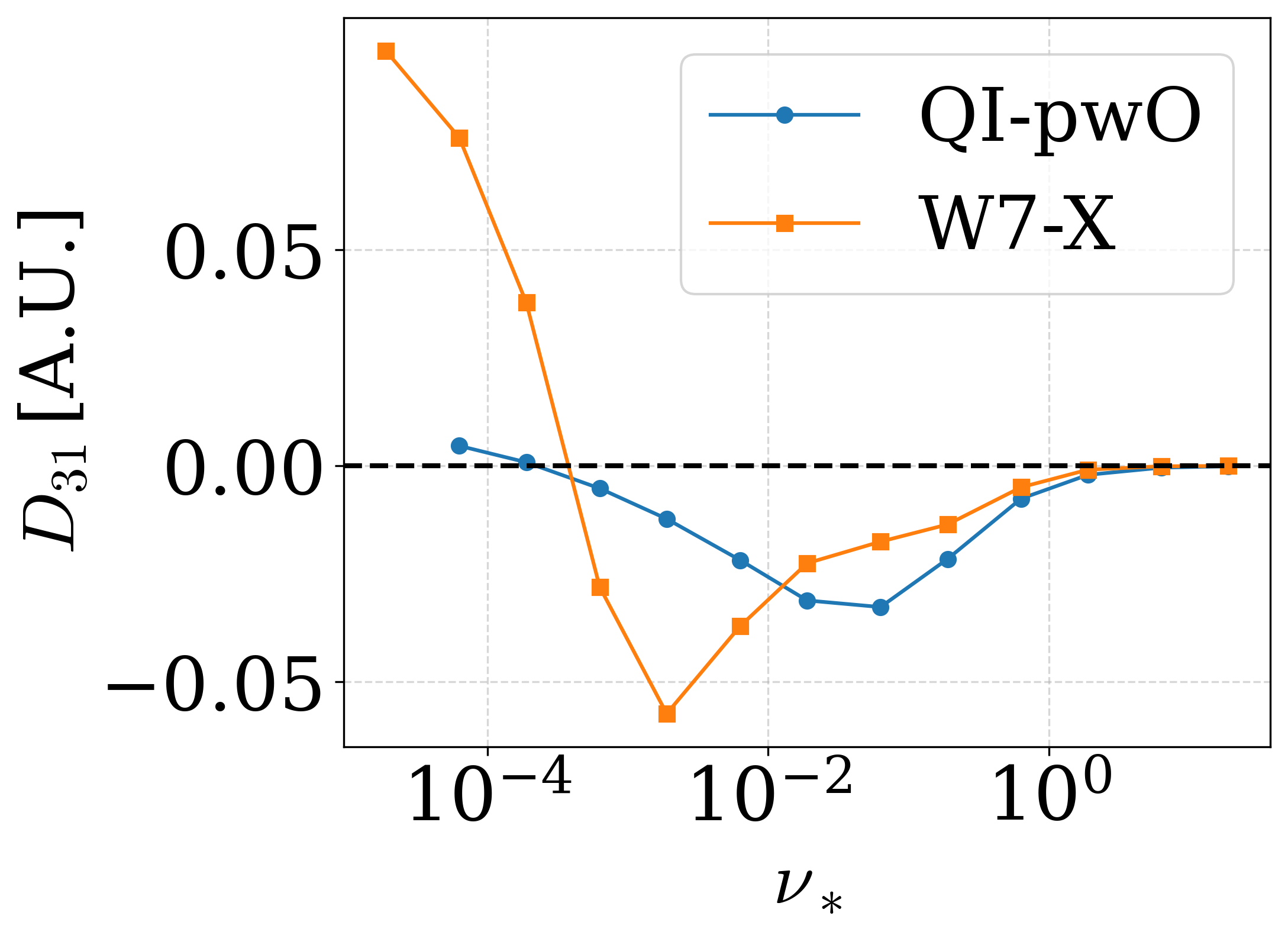}
\caption{Radial (left) and bootstrap (right) transport coefficients for the fields of figure~\ref{FIG_MODEL}.\label{FIG_D31}}
\end{figure}

In order to show this, in this section (and in the following sections, except otherwise stated) we use $N_{fp}=5$, $\iota=0.8815$, $B_{max}=3.0\,$T, $B_\mathrm{min}=2.1\,$T (which in turn yields $B_0=2.55\,$T and $\epsilon=0.45$), $\Delta\zeta=0.4$, $t_1=-0.08$, $(N_{fp}\zeta_c,\theta_c)=(\pi,\pi)$ and $p=4$. With these choices, the magnetic field resembles (with a broader buffer region) that of the high-mirror configuration of W7-X at $s=0.25$, as it can be seen in figure~\ref{FIG_MODEL} (here, $0 \le s \le1$ is a radial coordinate proportional to the toroidal flux). Figure~\ref{FIG_D31} shows the level of radial and parallel neoclassical transport, expressed in terms of the $D_{11}$ and $D_{31}$ transport coefficients, respectively, in the absence of a radial electric field (see e.g.~\cite{beidler2011ICNTS}), computed with~\texttt{MONKES}~\cite{escoto2024monkes}. For the model QI-pwO field, they are much smaller at low collisionality $\nu_*$. We note that $D_{31}$ is expected to deviate from zero at low enough $\nu_*$, coinciding with the appearance of a $1/\nu$ regime for $D_{11}$, since the analytical proof of zero bootstrap current requires tokamak-like banana transport~\cite{calvo2025pwO}.

\section{Parameter scan}\label{SEC_SCANS}

In this section, we perform scans in the different parameters of the model field of equation (\ref{EQ_QIPWO}). In all the cases that will be shown, the radial neoclassical transport is negligible, as in figure~\ref{FIG_D31} (left). In all cases, except for some during the scan in $w_2$, the bootstrap current will also be small, as in figure~\ref{FIG_D31} (right). These parameter scans will then provide a better understanding of the different shapes of the contours of minimum $B$ (and consequently different shapes of the pwO region) that can be made consistent with reduced neoclassical radial transport  and bootstrap current in a QI-pwO field.

\begin{figure}
\includegraphics[angle=0,width=.49\columnwidth]{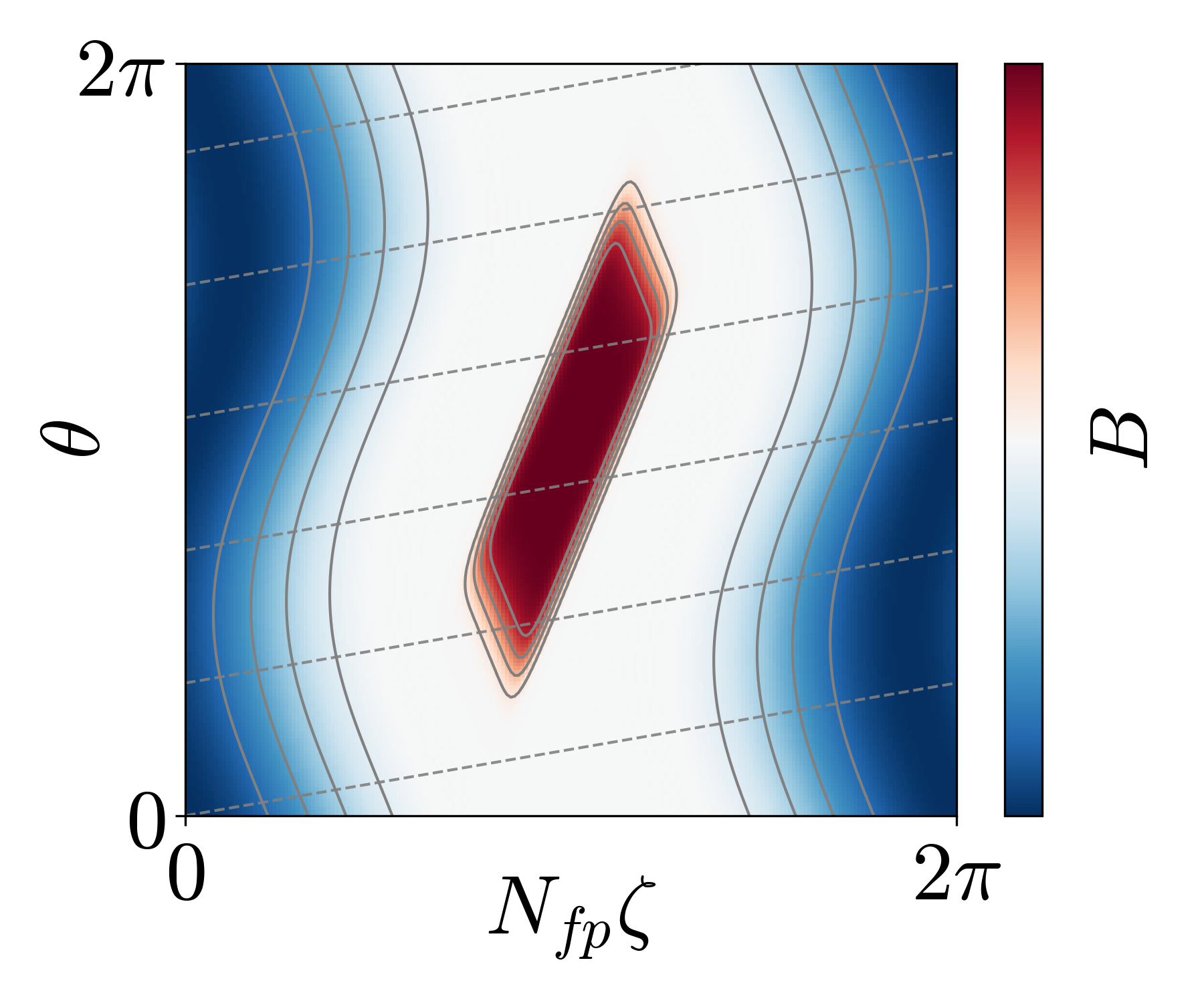}  
\includegraphics[angle=0,width=.49\columnwidth]{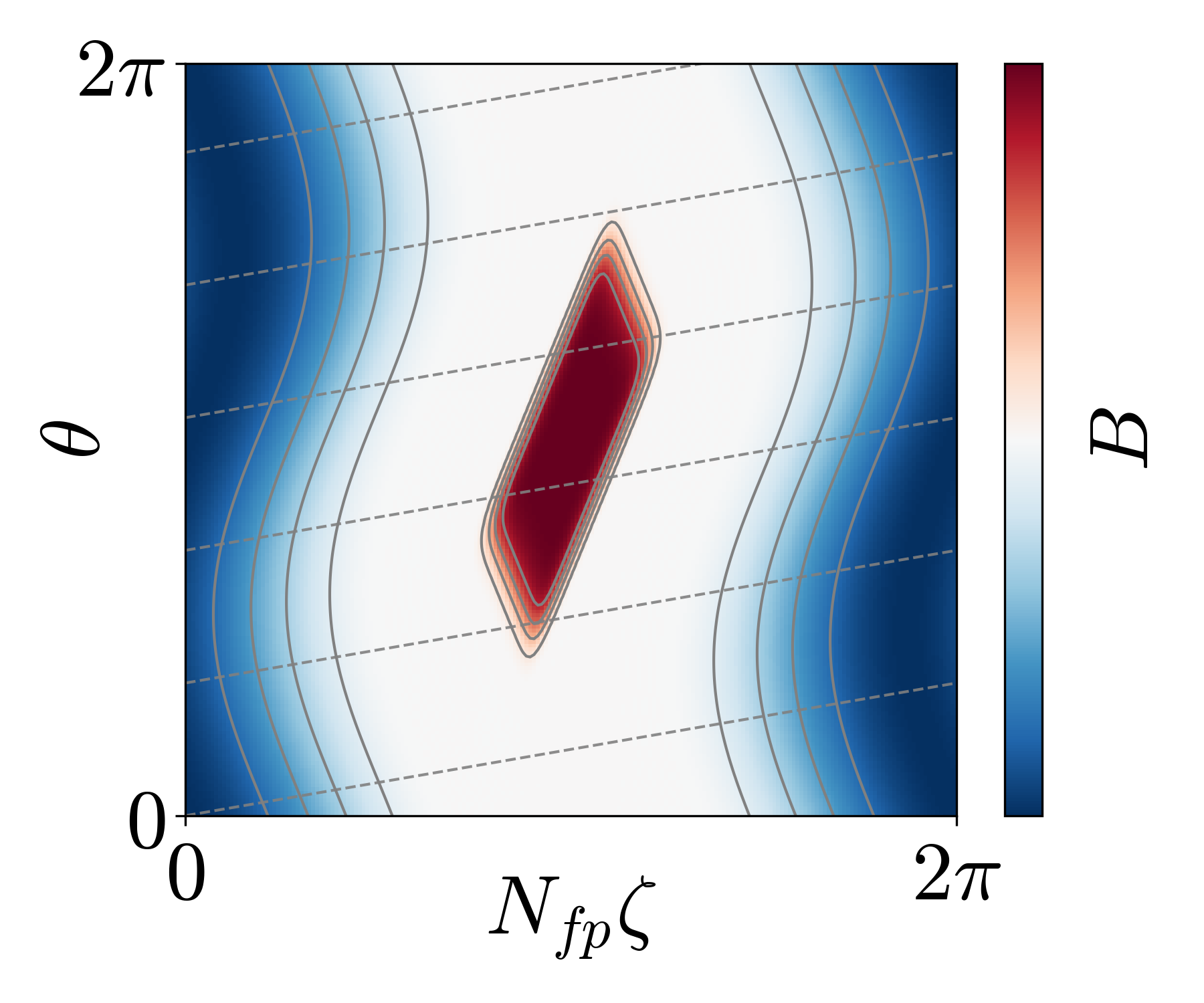}
\includegraphics[angle=0,width=.49\columnwidth]{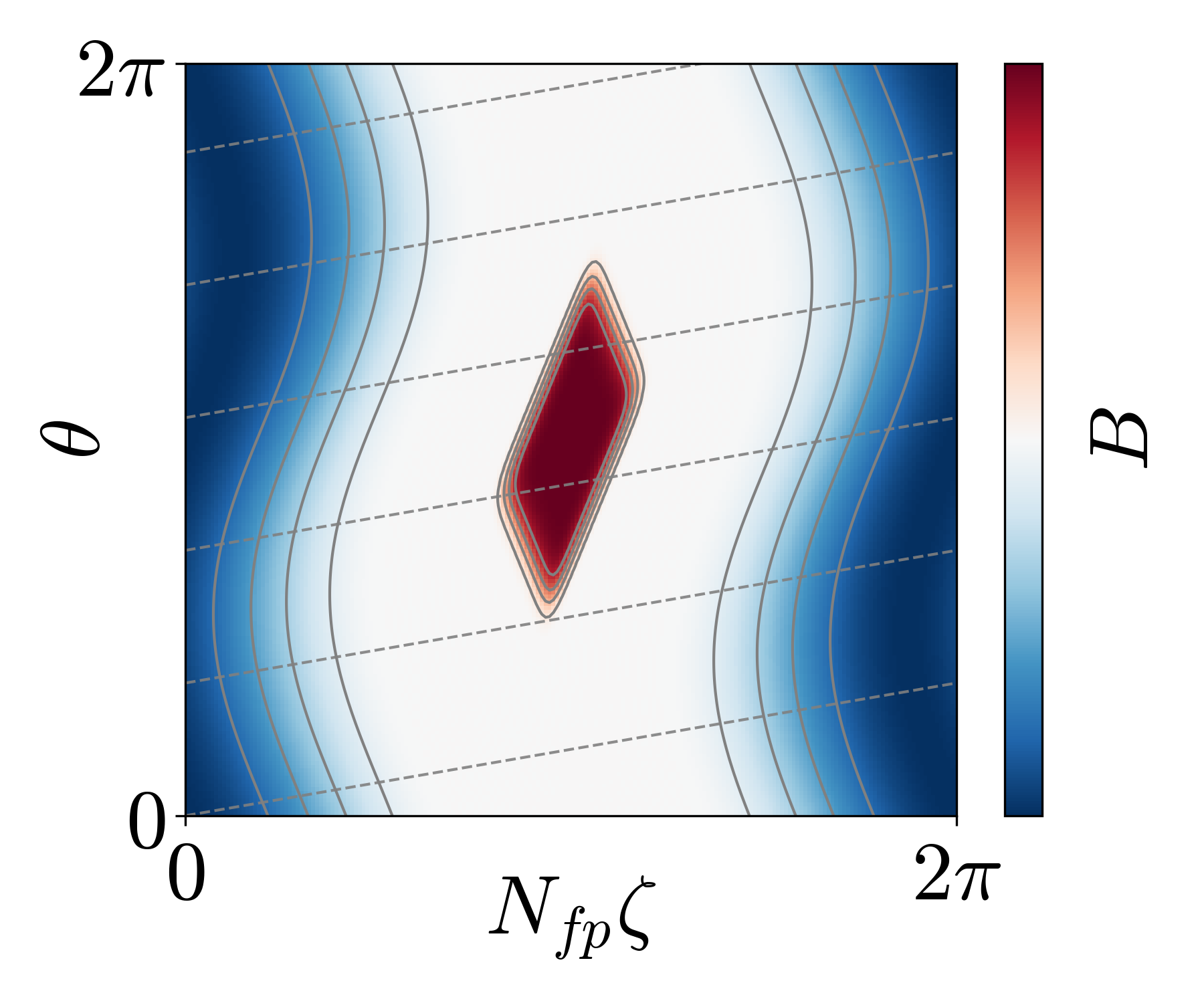}
\includegraphics[angle=0,width=.49\columnwidth]{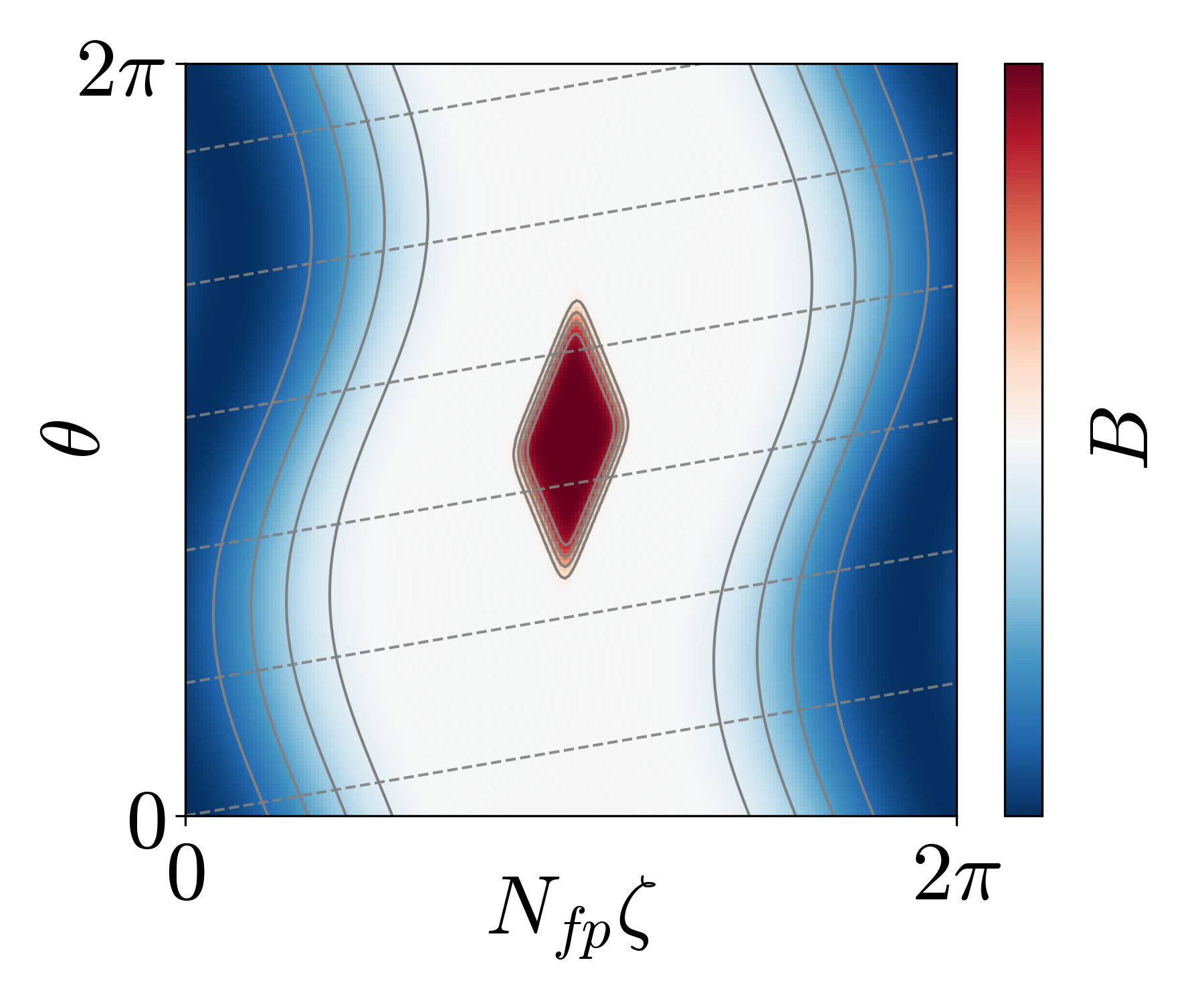}
\caption{Scan in $w_2$: from left to right and top to bottom, $w_2=\pi$, $w_2=0.8\pi$, $w_2=0.6\pi$, and $w_2=0.4\pi$.\label{FIG_W2}}
\end{figure}

\begin{figure}
\includegraphics[angle=0,width=.49\columnwidth]{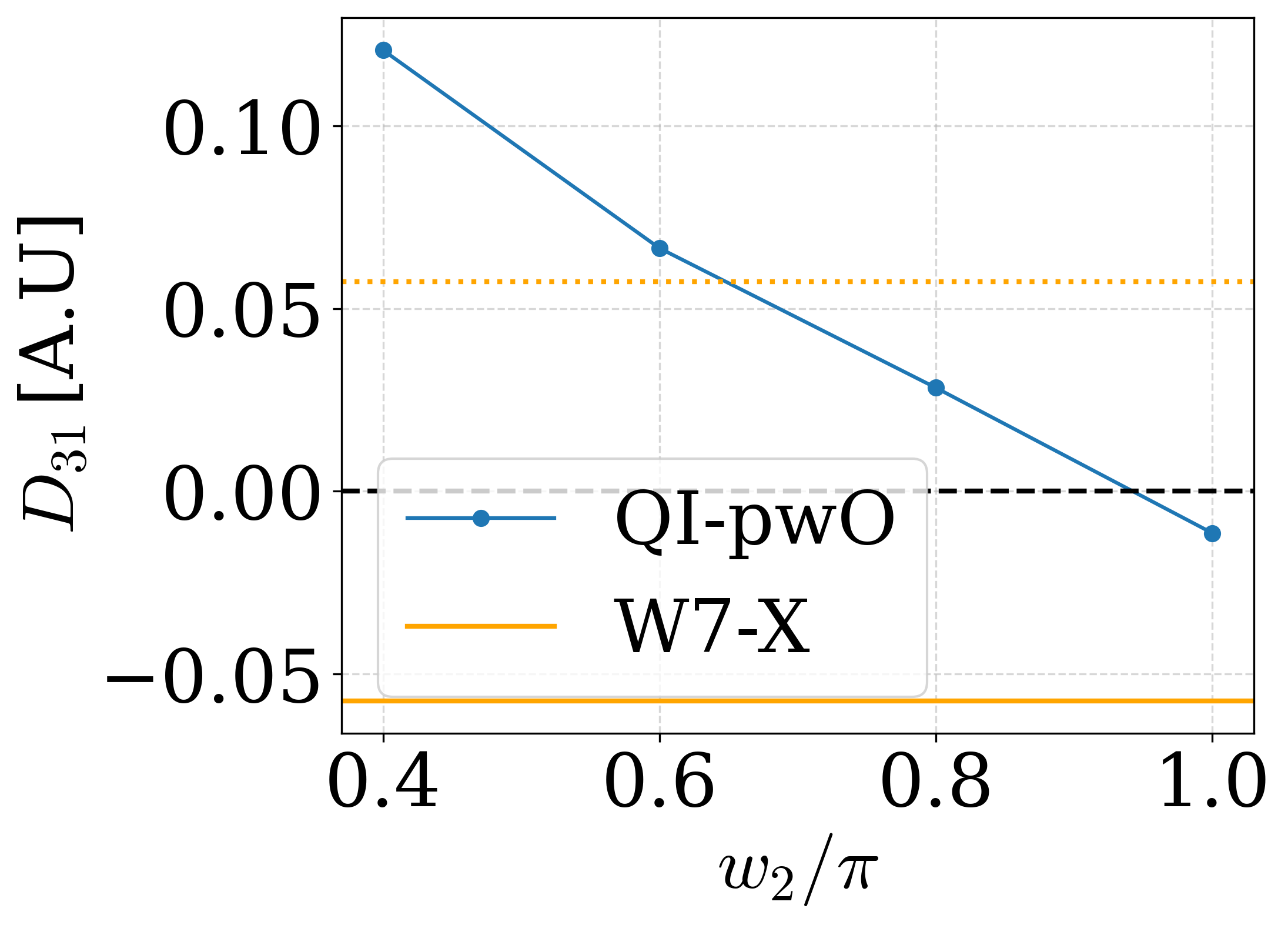}  
\caption{Bootstrap transport coefficient as a function of $w_2$ for the scan of figure~\ref{FIG_W2} (orange: $\pm D_{31}$ for W7-X).\label{FIG_D31W2}}
\end{figure}

In section~\ref{SEC_MODEL}, the model QI-pwO field was both closer to being QI (in the region $B\gtrsim B_\mathrm{min}$) and pwO (in the region $B\lesssim B_\mathrm{max}$) than W7-X. One then cannot rule out, in principle, that closeness to quasi-isodynamicity around $B\gtrsim B_\mathrm{min}$ was the responsible feature for the reduced bootstrap current with respect to W7-X. In order to rule this out, we perform a scan in $w_2$ while keeping the rest of the parameters constant, see figure~\ref{FIG_W2} (we set $t_1=-0.04$, which allows for a wider scan). Figure \ref{FIG_D31W2} then shows the $D_{31}$ coefficient for $\nu_*=2\times 10^{-3}$. It is significantly closer to zero than W7-X high-mirror only in an interval around $w_2\approx \pi$, which confirms that the shape of the pwO deviation from quasi-isodynamicity matters for parallel neoclassical transport.

\begin{figure}[h!]
\includegraphics[angle=0,width=.49\columnwidth]{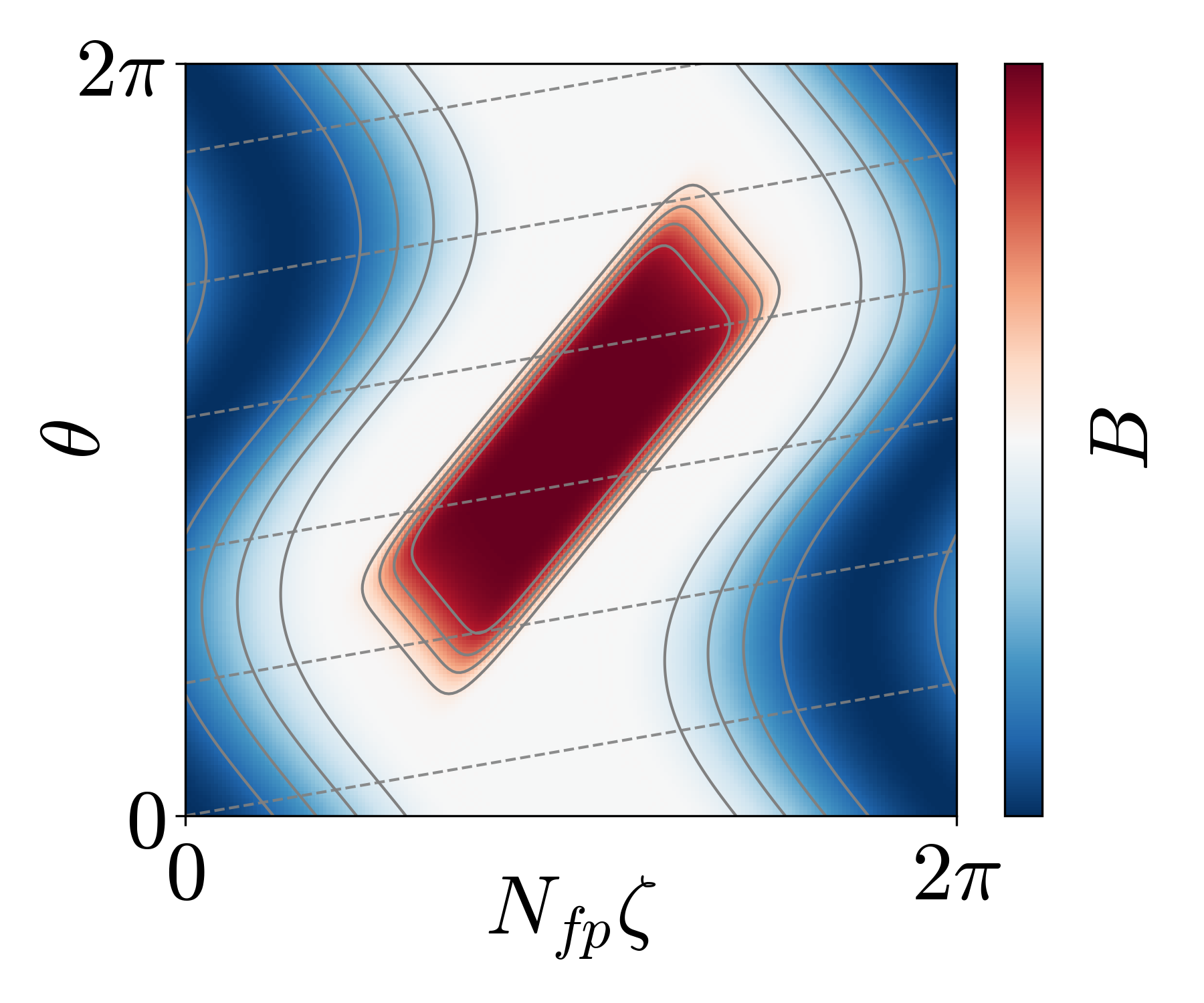}  
\includegraphics[angle=0,width=.49\columnwidth]{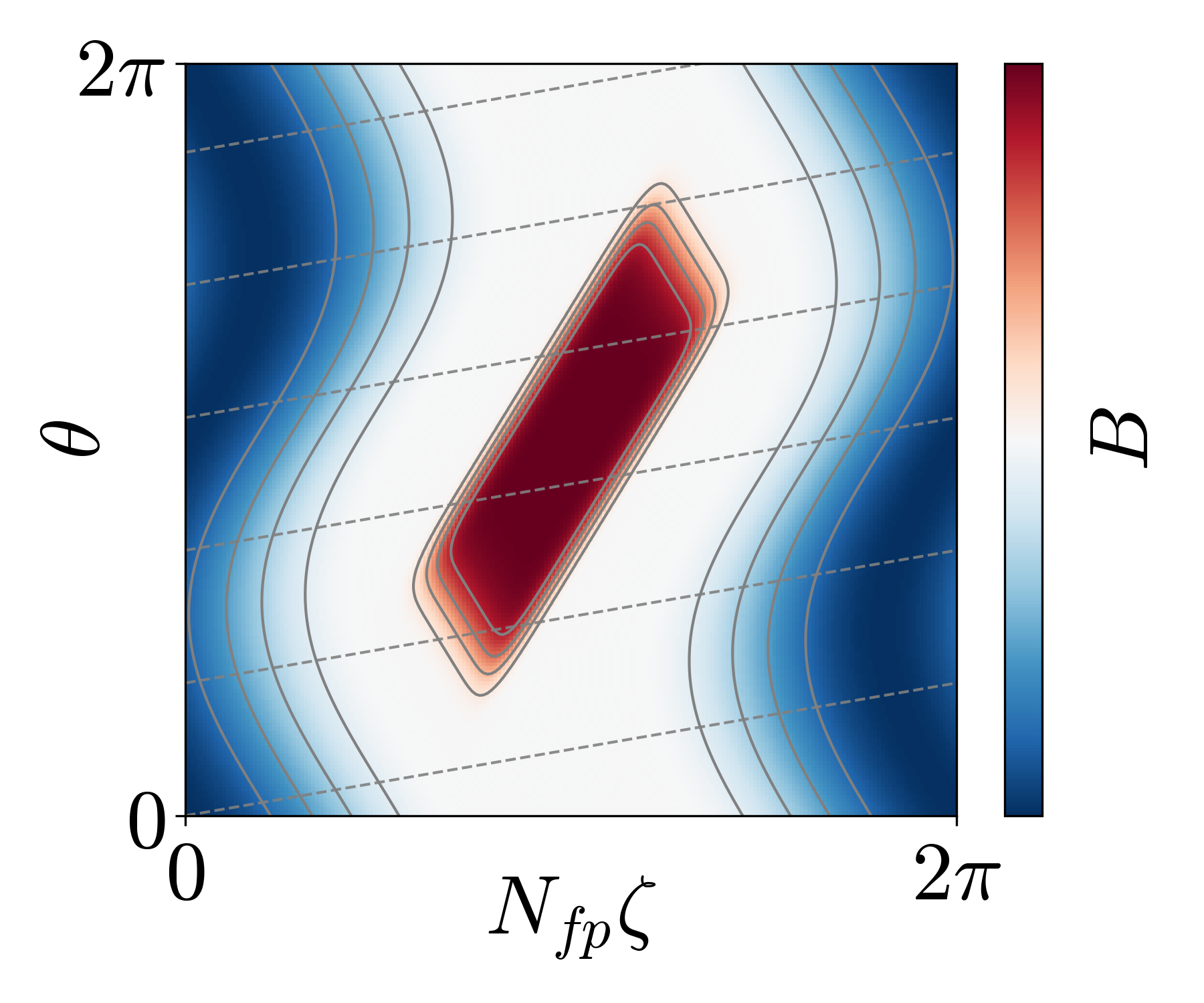}  
\includegraphics[angle=0,width=.49\columnwidth]{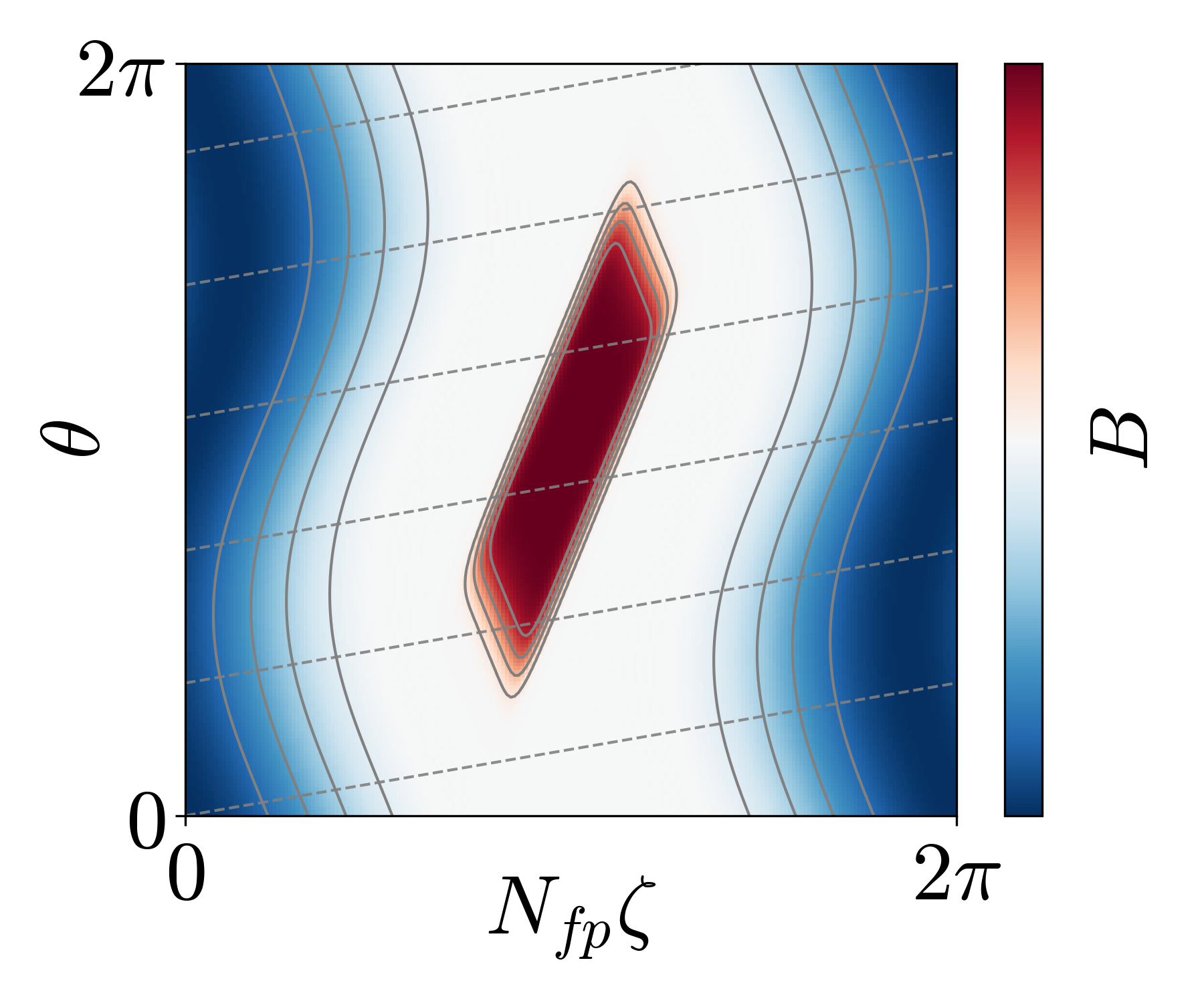}  
\includegraphics[angle=0,width=.49\columnwidth]{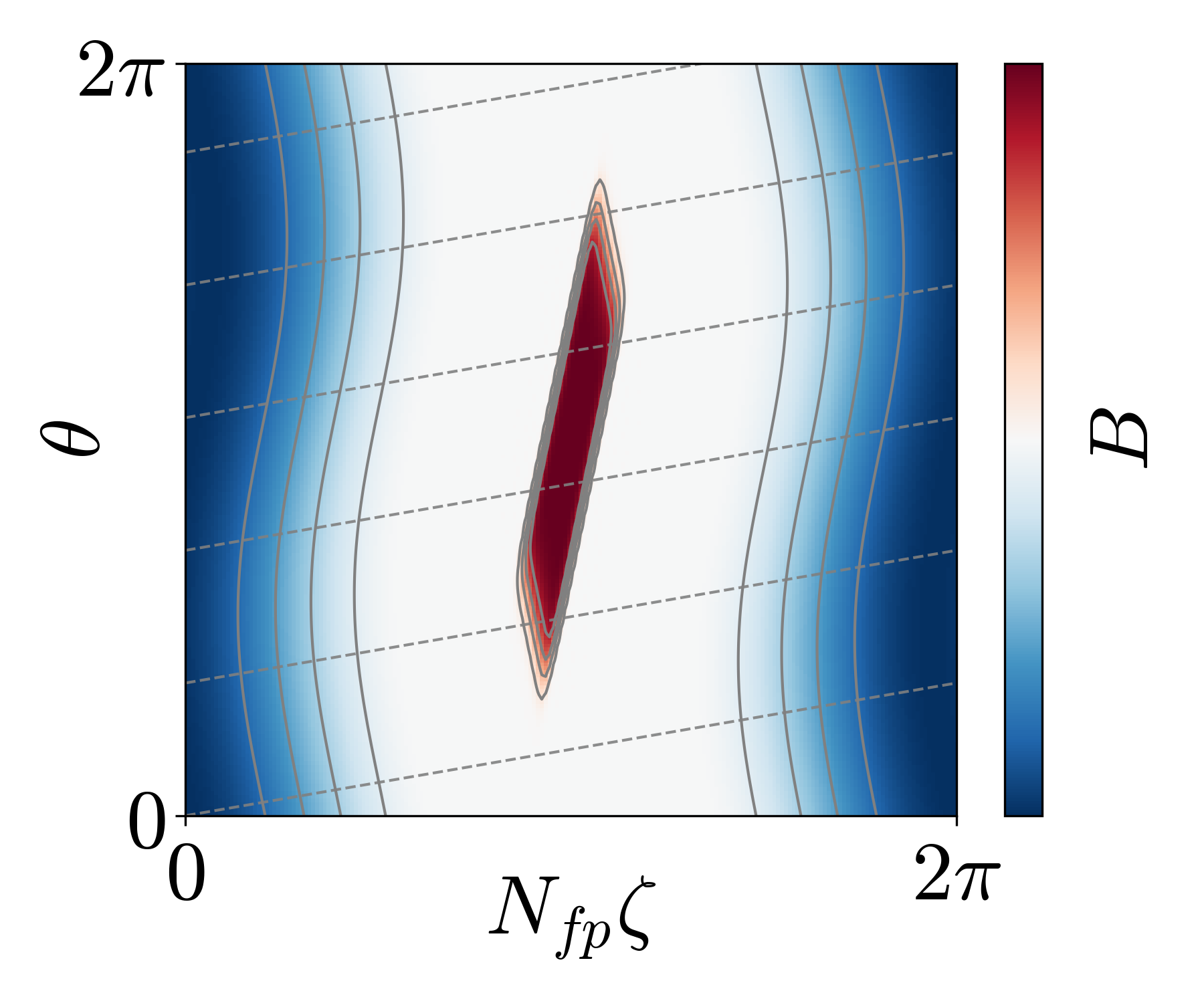}  
\caption{Scan in $t_1$: from left to right and top to bottom, $t_1=-0.16$, $t_1=-0.12$, $t_1=-0.08$, and $t_1=-0.04$. \label{FIG_T1}}
\end{figure}

In figure \ref{FIG_T1}, we perform a scan in $t_1$ while keeping the rest of the parameters constant (except for $w_1$ and $t_2$, which change according to equations (\ref{EQ_W1}) and (\ref{EQ_T2})).  For values of $t_1$ closer to zero, the $B=B_\mathrm{max}$ region of the flux surface becomes narrower and, at the same time, the $B$-contours QI region of the field are becoming independent of $\zeta$. In the limit of $t_1=0$, $B$ becomes QI (especifically, quasi-poloidally symmetric). That QI-pwO fields can be smoothly connected to QI fields (although, numerically, the construction of equation~(\ref{EQ_QIPWO}) may need to be slightly modified to ensure periodicity, see~\cite{escoto2025qi}) is encouraging for the prospect of using the concept of piecewise omnigenity to modify QI configurations to make them more suitable for a stellarator reactor.

\begin{figure}
\includegraphics[angle=0,width=.49\columnwidth]{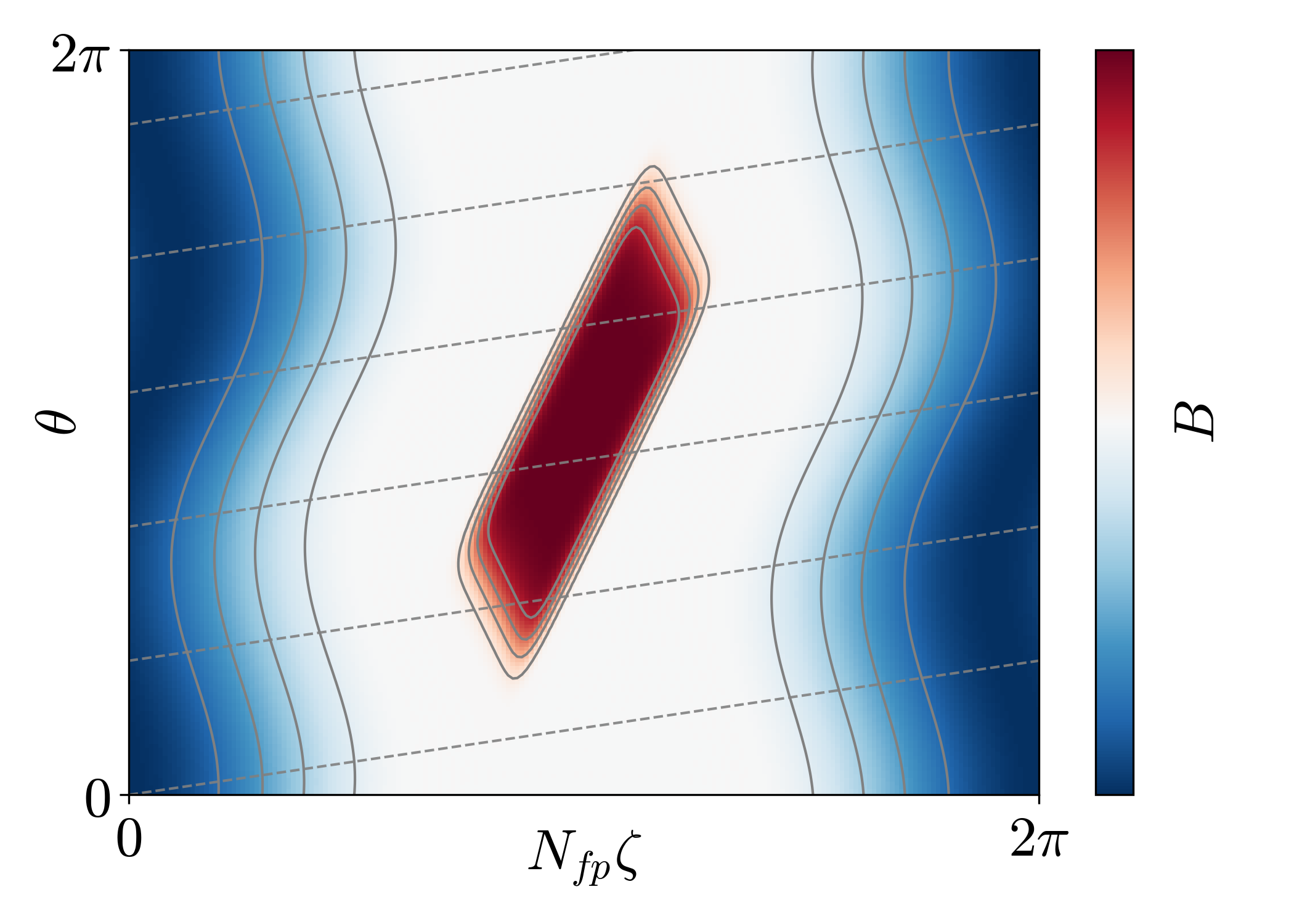}  
\includegraphics[angle=0,width=.49\columnwidth]{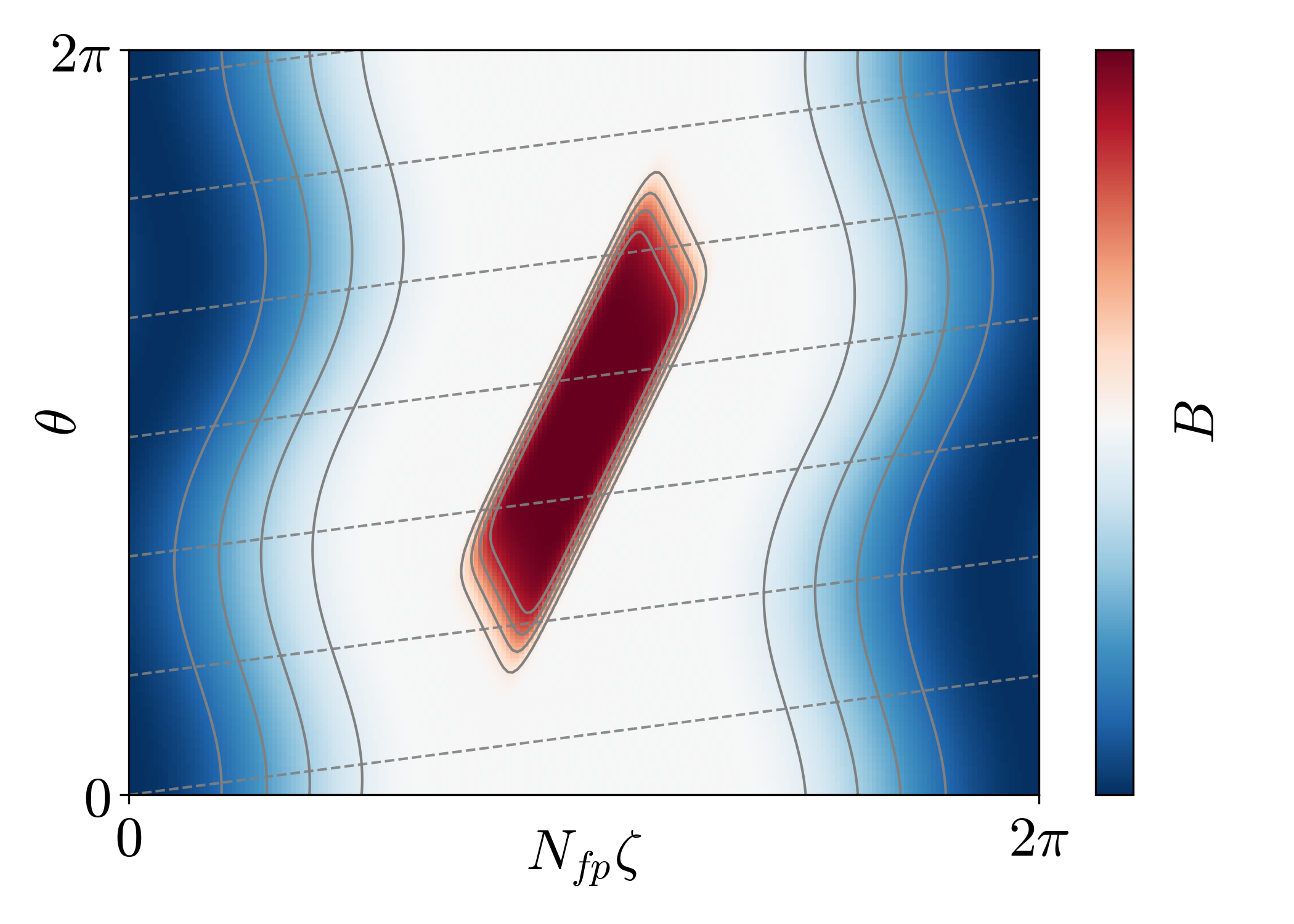}
\includegraphics[angle=0,width=.49\columnwidth]{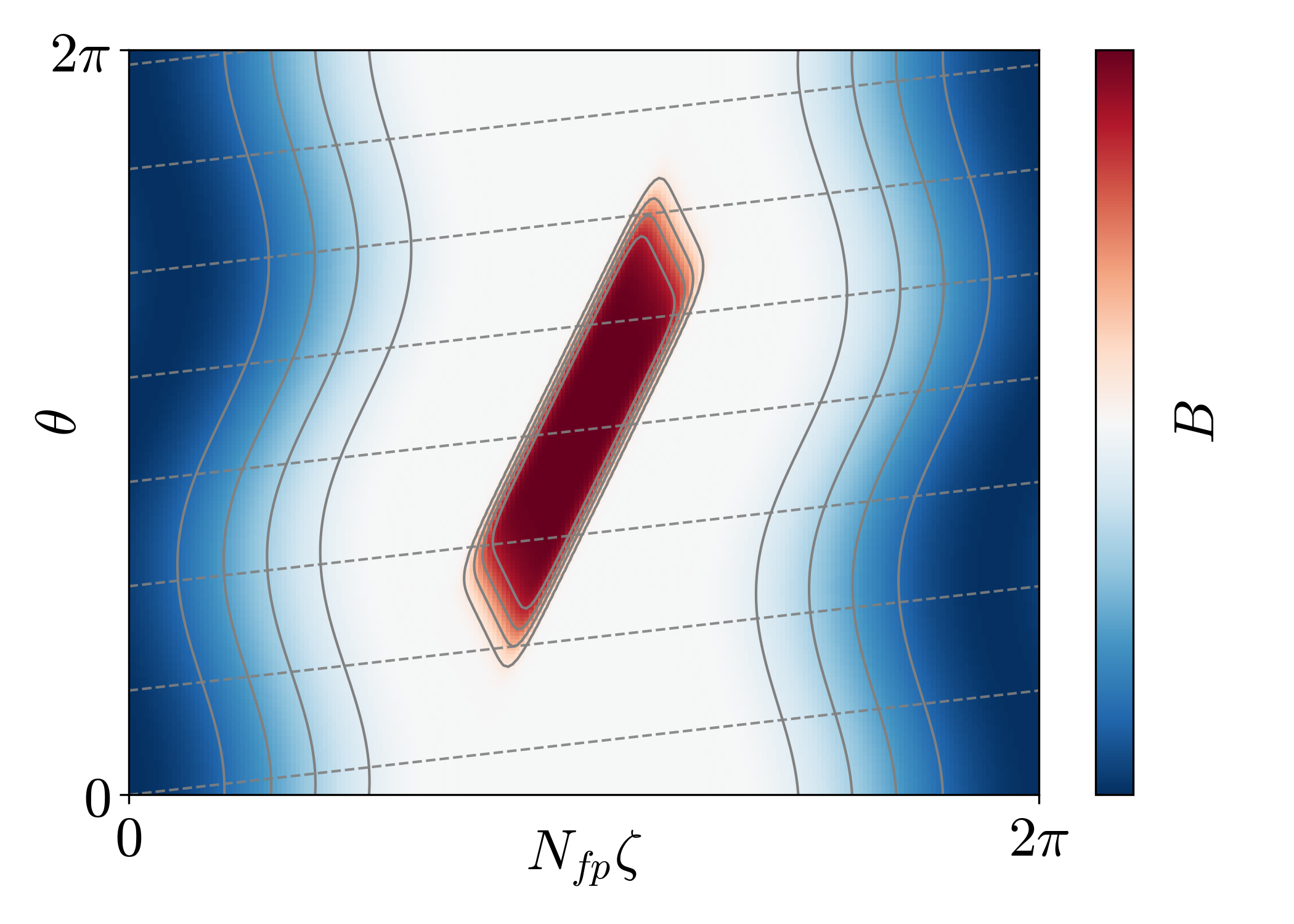}
\includegraphics[angle=0,width=.49\columnwidth]{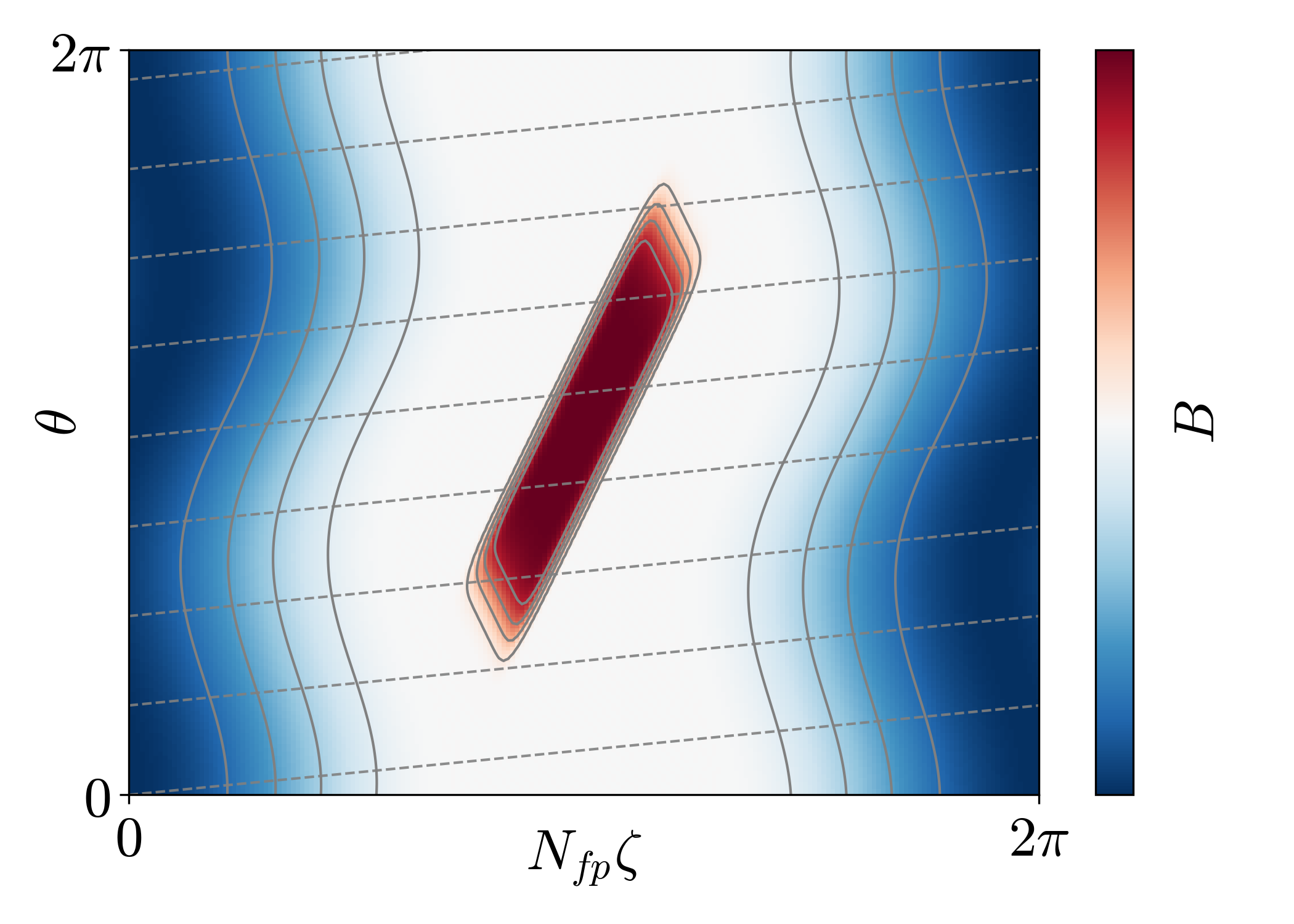}  
\caption{Scan in $\iota$ at constant $t_1$: from left to right and top to bottom, $\iota=0.9$, $\iota=0.8$, $\iota=0.7$, and $\iota=0.6$.\label{FIG_IOTA}}
\end{figure}


%
%

Periodicity has an impact on the allowable deviations from quasi-isodynamicity. It is straightforward to see that the results presented in this work (and, specifically, figures~\ref{FIG_W2} and \ref{FIG_T1}) do not change under the transformation
\begin{eqnarray}
N_{fp} &\rightarrow& kN_{fp}\,,\\
B(\theta,\zeta) &\rightarrow& B(\theta,k\zeta)\,,\\
\iota &\rightarrow& k\iota \,,\\N_{fp}
t_1 &\rightarrow& k^{-1}t_1\,,\\
t_2 &\rightarrow& kt_2\,,\\
w_1 &\rightarrow& k^{-1}w_1\,,\\
\Delta\zeta &\rightarrow& k^{-1}\Delta\zeta\,,
\end{eqnarray}
where $k$ is a natural number. For instance, a QI-pwO field with $N_{fp}=10$ and $B(N_{fp}\zeta,\theta)$ similar to W7-X (as in figure~\ref{FIG_MODEL}) would require a value of $\iota$ close to 2, which would likely complicate the shape of a configuration with this spatial variation of $B$. The equivalent QI-pwO field with $N_{fp}=1$ would require a value of $\iota$ below 0.2, probably too small to provide MHD stability.

\begin{figure}
\includegraphics[angle=0,width=.49\columnwidth]{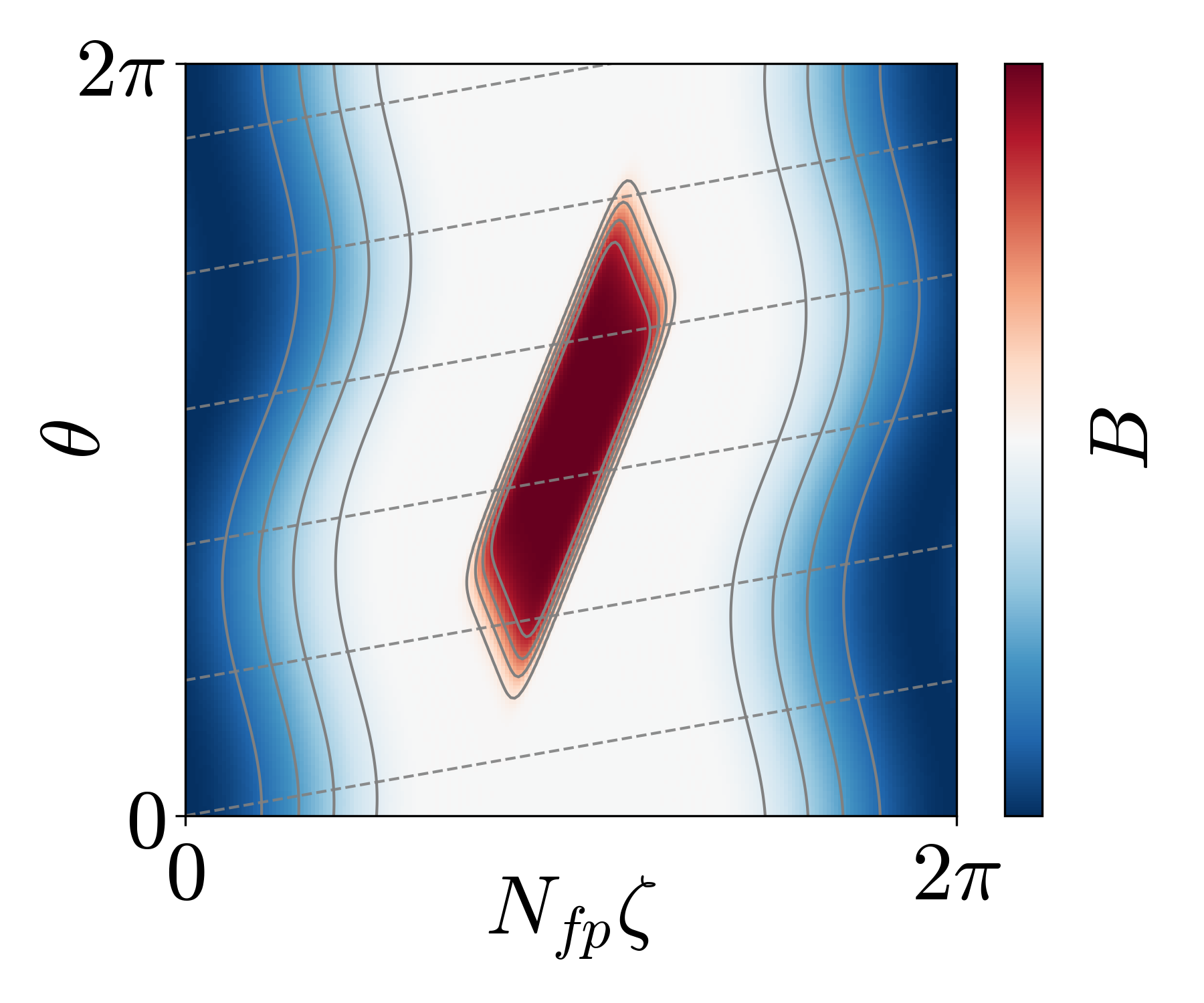}  
\includegraphics[angle=0,width=.49\columnwidth]{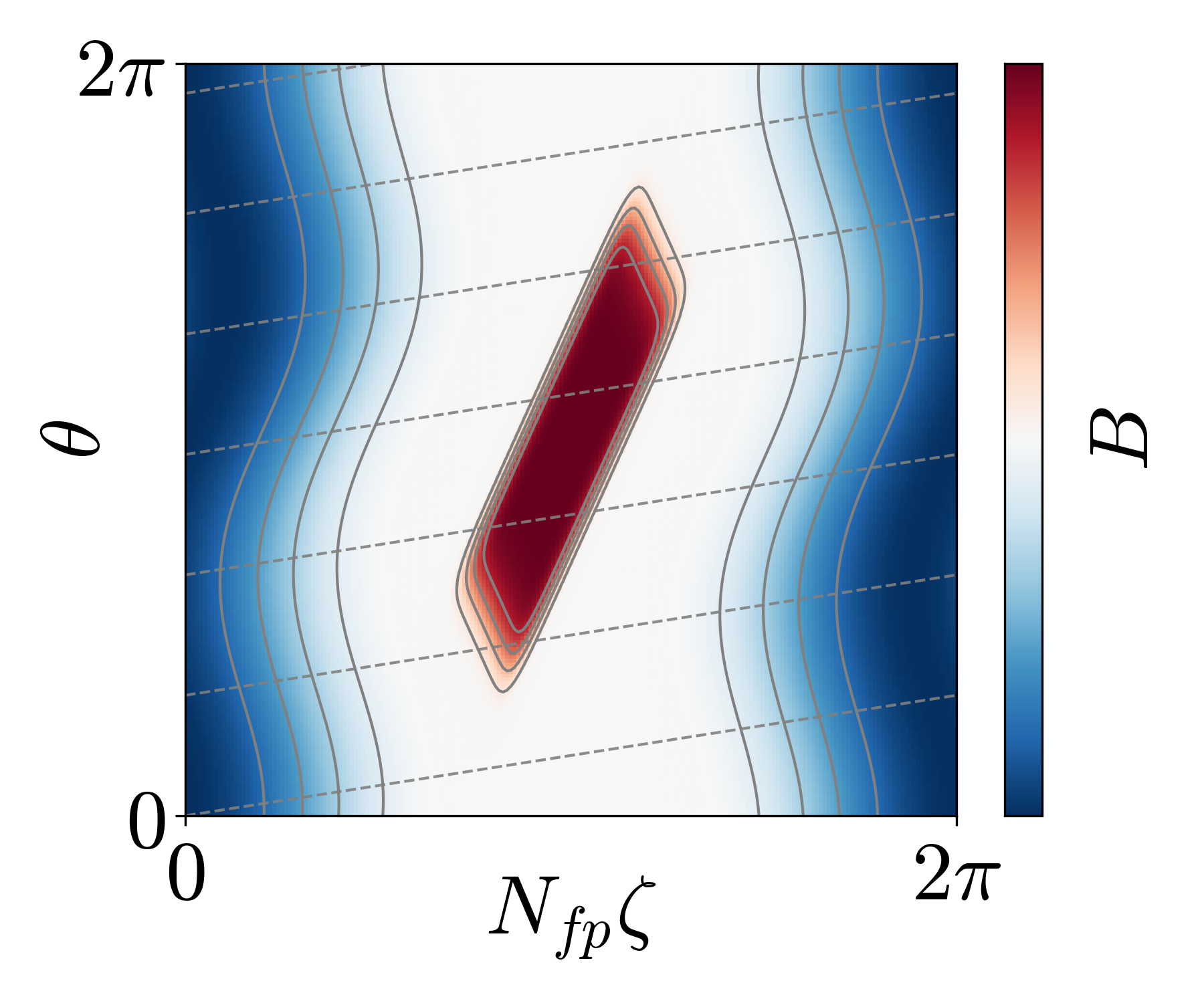}
\includegraphics[angle=0,width=.49\columnwidth]{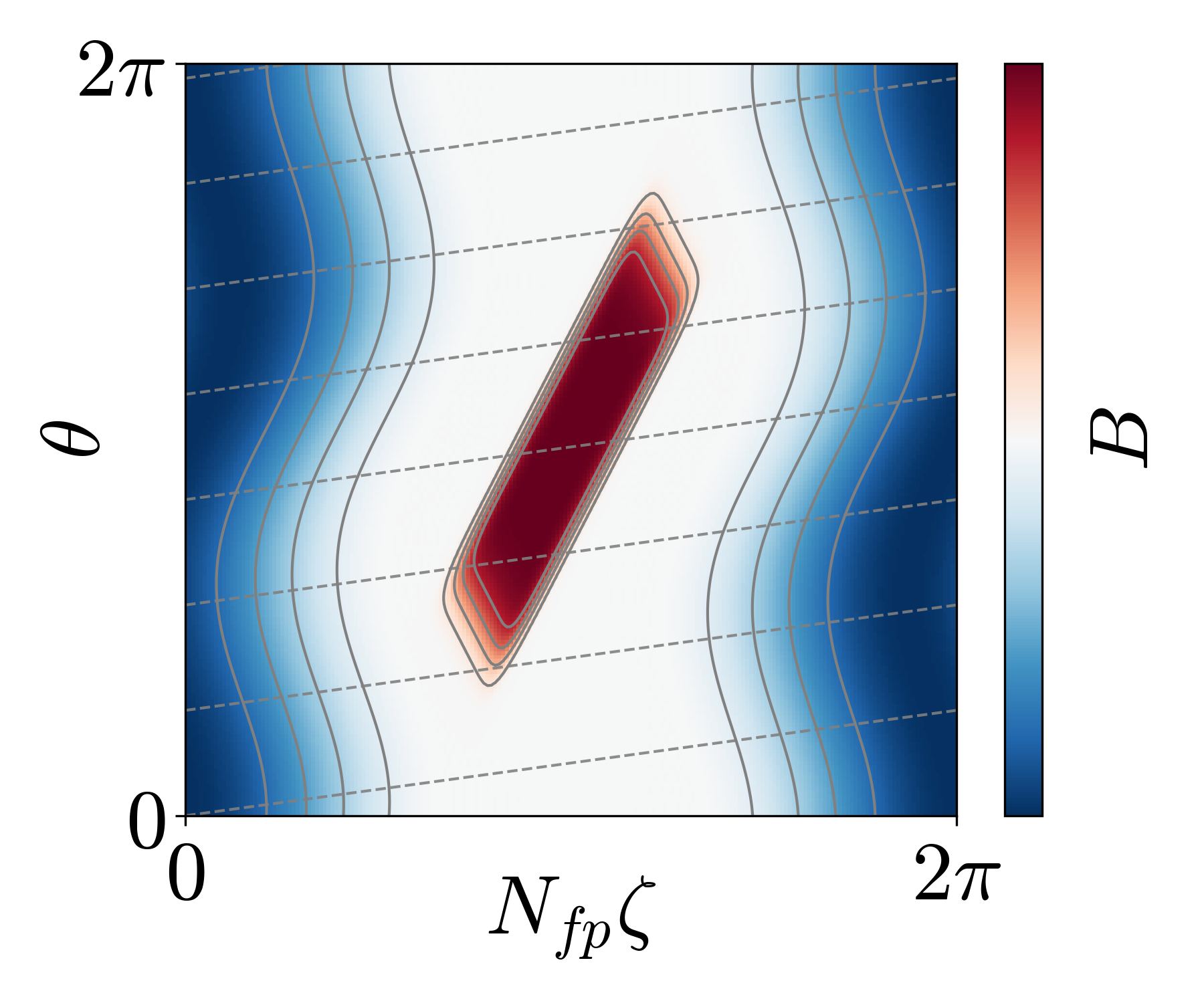}
\includegraphics[angle=0,width=.49\columnwidth]{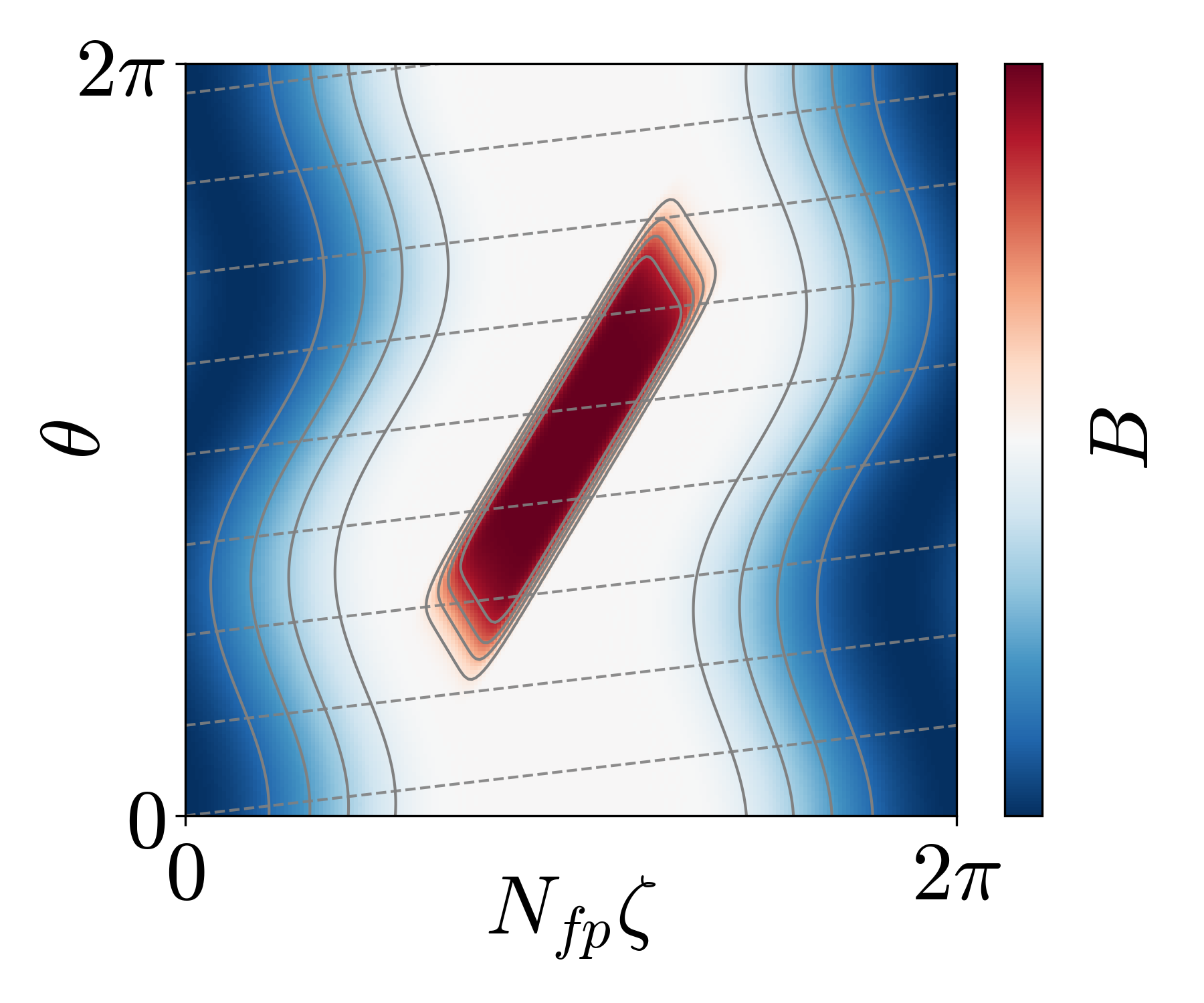}  
\caption{Scan in $\iota$ at constant $w_1$: from left to right and top to bottom, $\iota=0.9$, $\iota=0.8$, $\iota=0.7$, and $\iota=0.6$.\label{FIG_IOTAB}}
\end{figure}

Finally, in such cases, one would need to modify $\iota$ further and change $w_1$ and/or $t_1$ accordingly. In figure \ref{FIG_IOTA}, we perform a scan in $\iota$ while keeping $t_1$ constant. In figure \ref{FIG_IOTAB}, we scan in $\iota$ while keeping $w_1$ constant. We see that smaller $\iota$, via smaller $w_1$ or larger $|t_1|$, leads to a narrower or more tilted pwO region, respectively

\begin{figure}
\includegraphics[angle=0,width=.49\columnwidth]{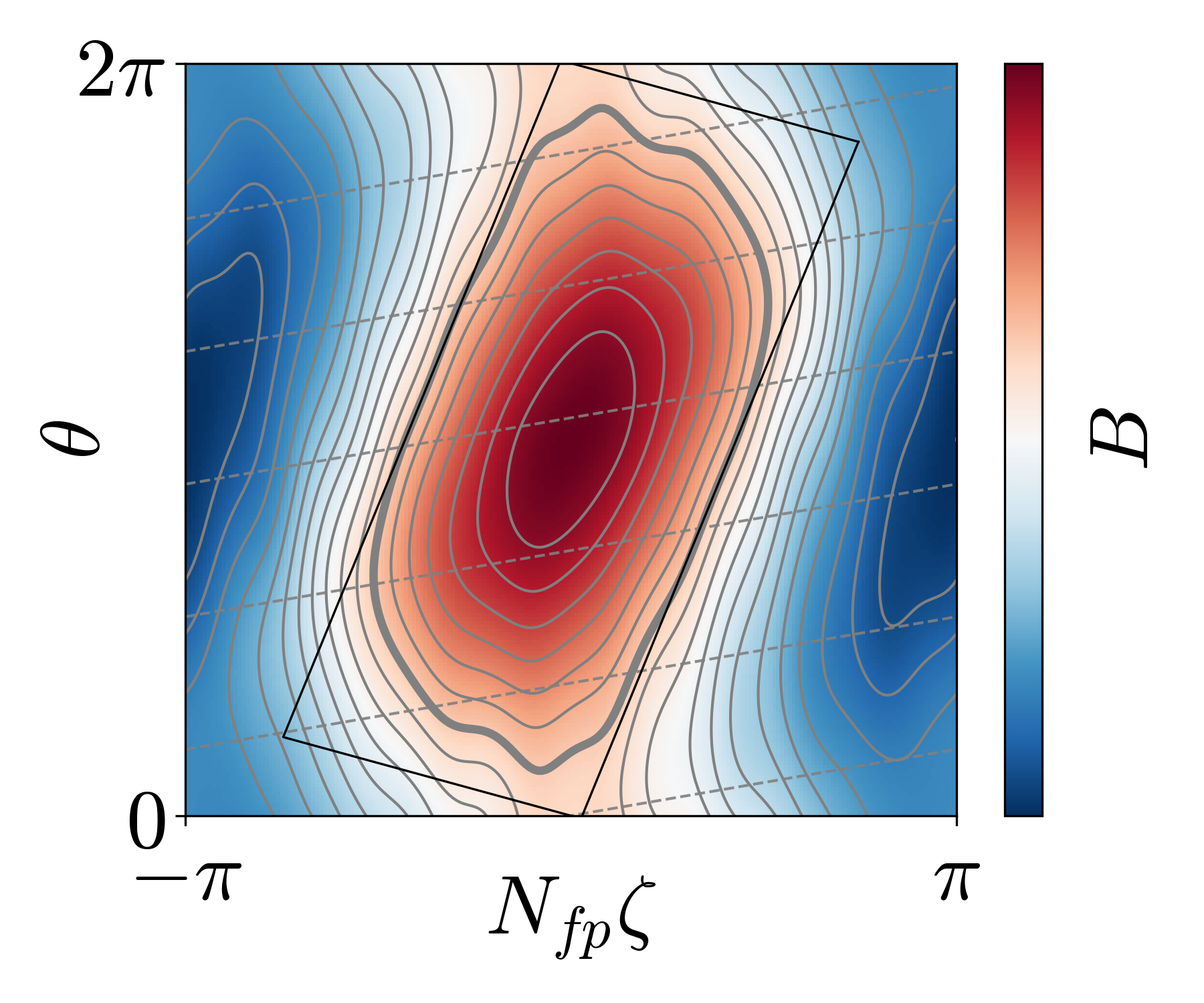}  
\includegraphics[angle=0,width=.49\columnwidth]{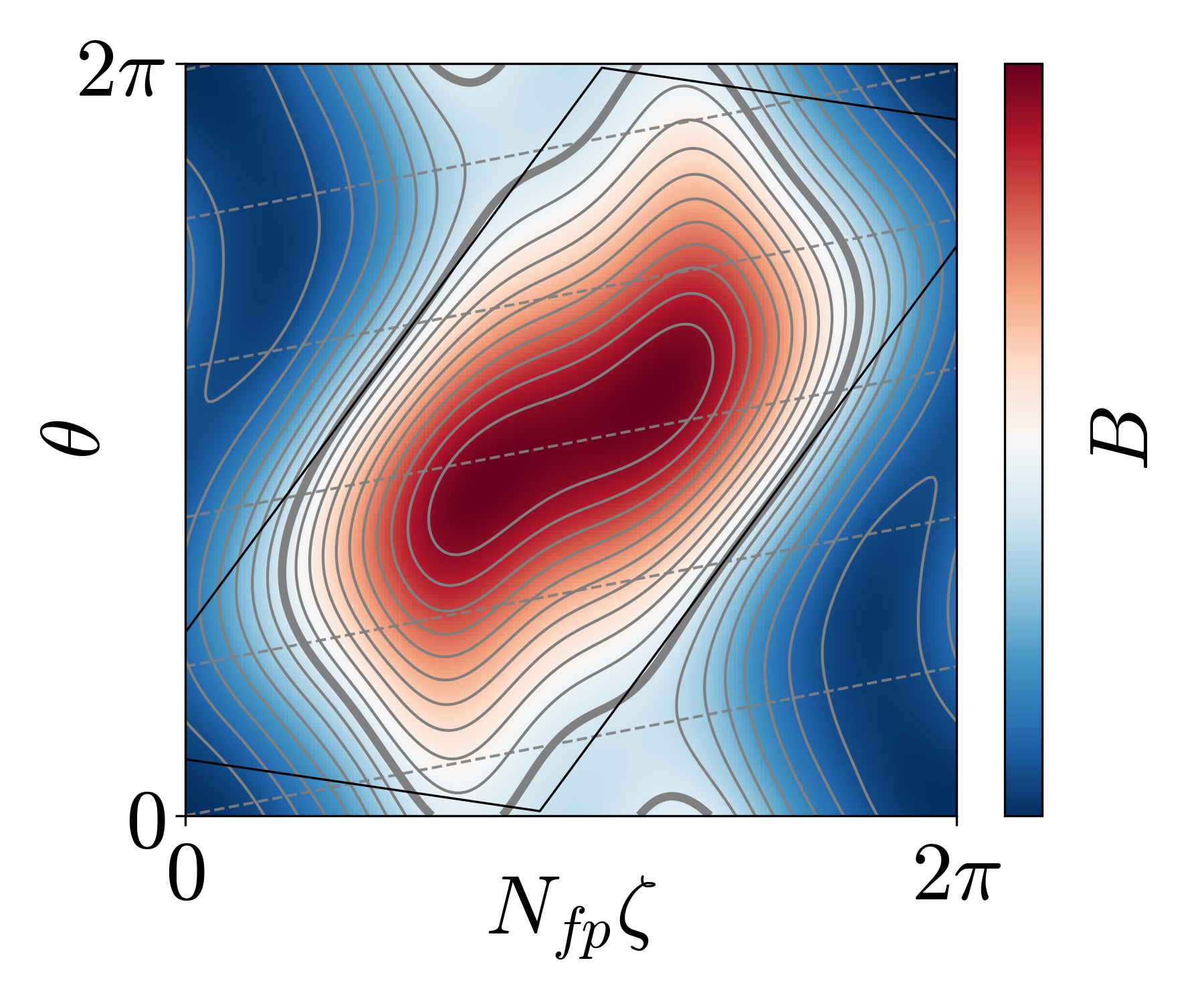}  
\includegraphics[angle=0,width=.49\columnwidth]{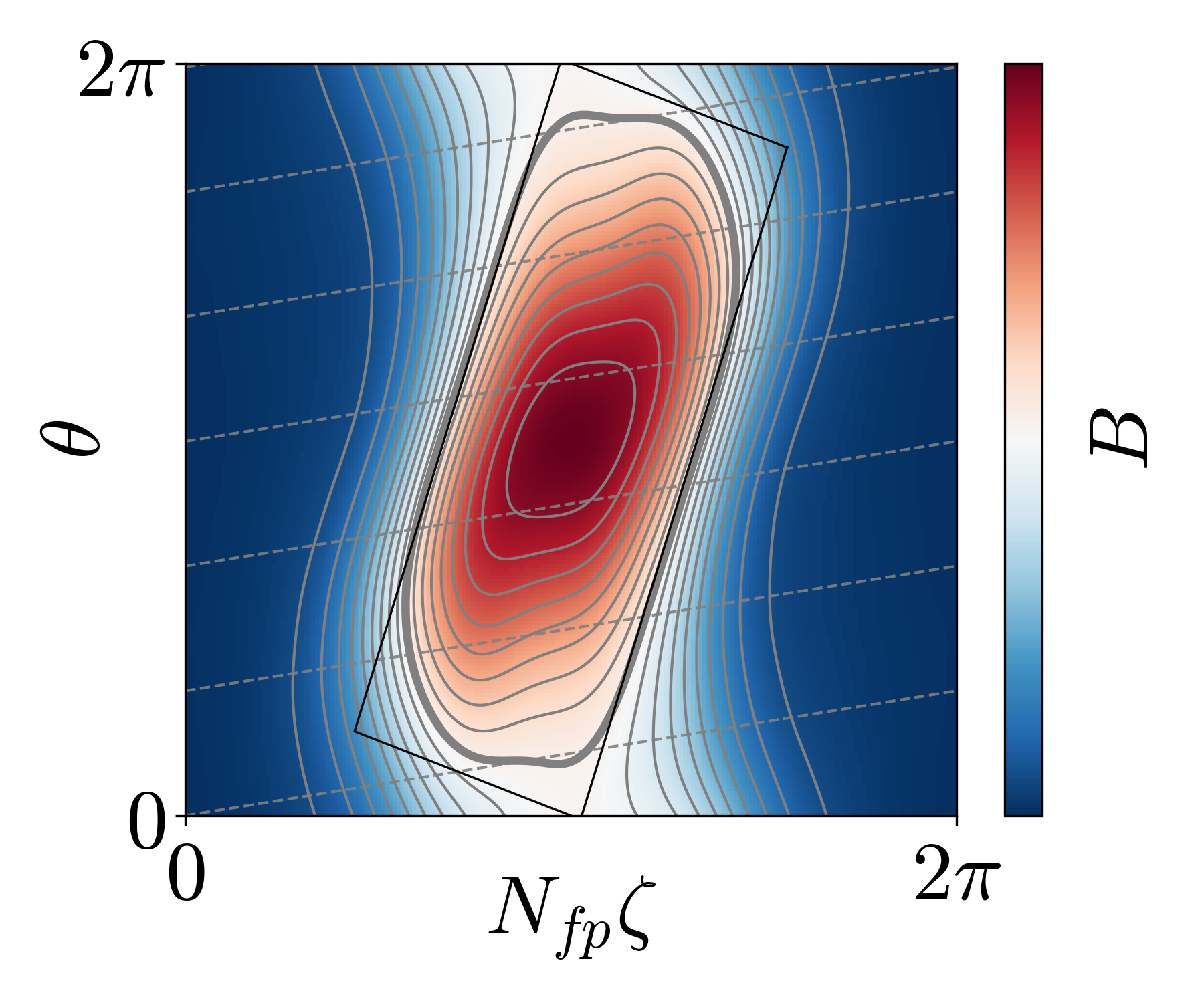}
\includegraphics[angle=0,width=.49\columnwidth]{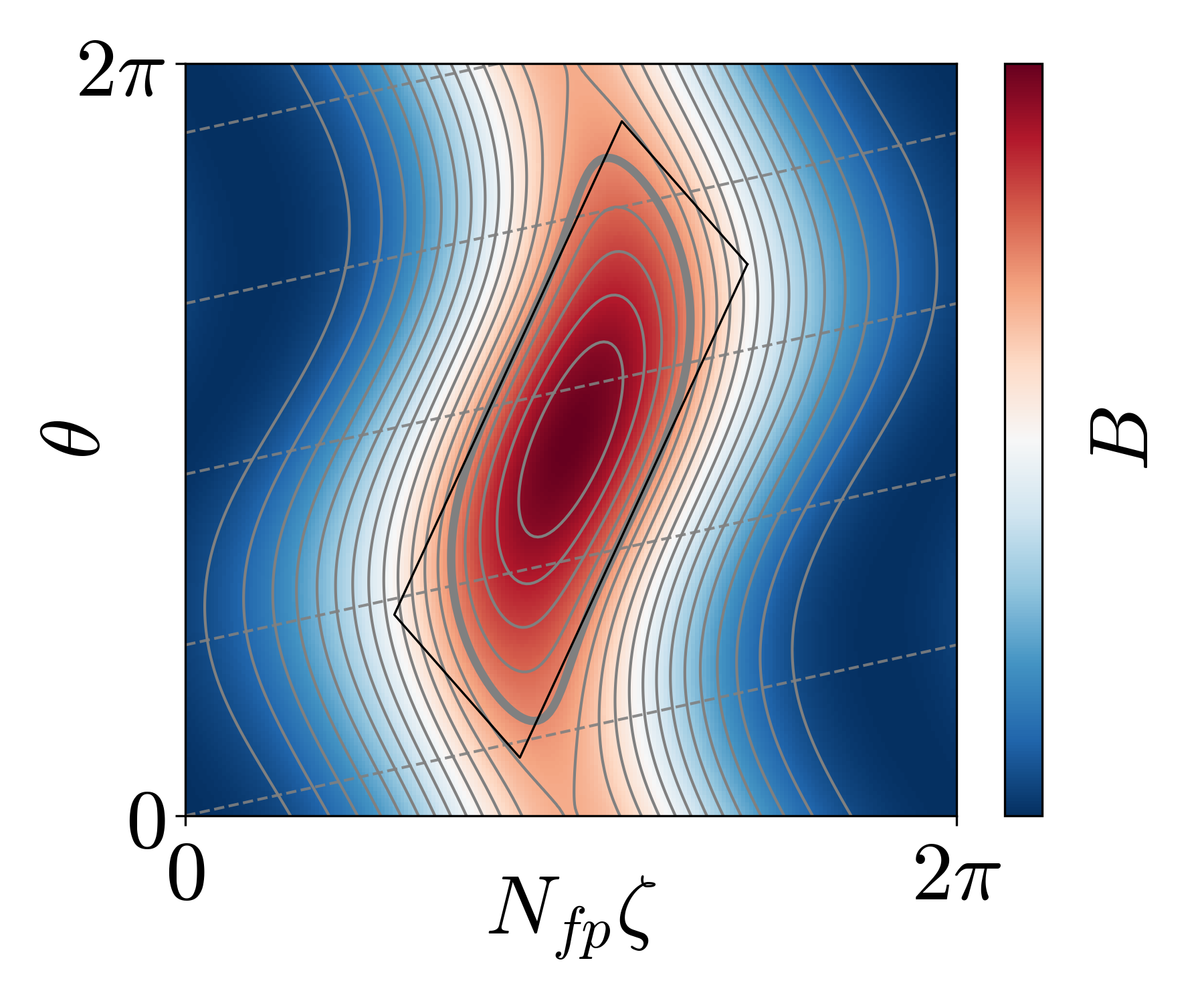}
\caption{Magnetic field strength of W7-X-high mirror (top, left), conf. A (top, right), CIEMAT-pw1 (bottom, left), and USTC-pwO3 (bottom, right). Overlaid: characteristic parallelogram of a pwO field with vanishing bootstrap current, see text. \label{FIG_CONFS}}
\end{figure}

\begin{figure}
\includegraphics[angle=0,width=.99\columnwidth]{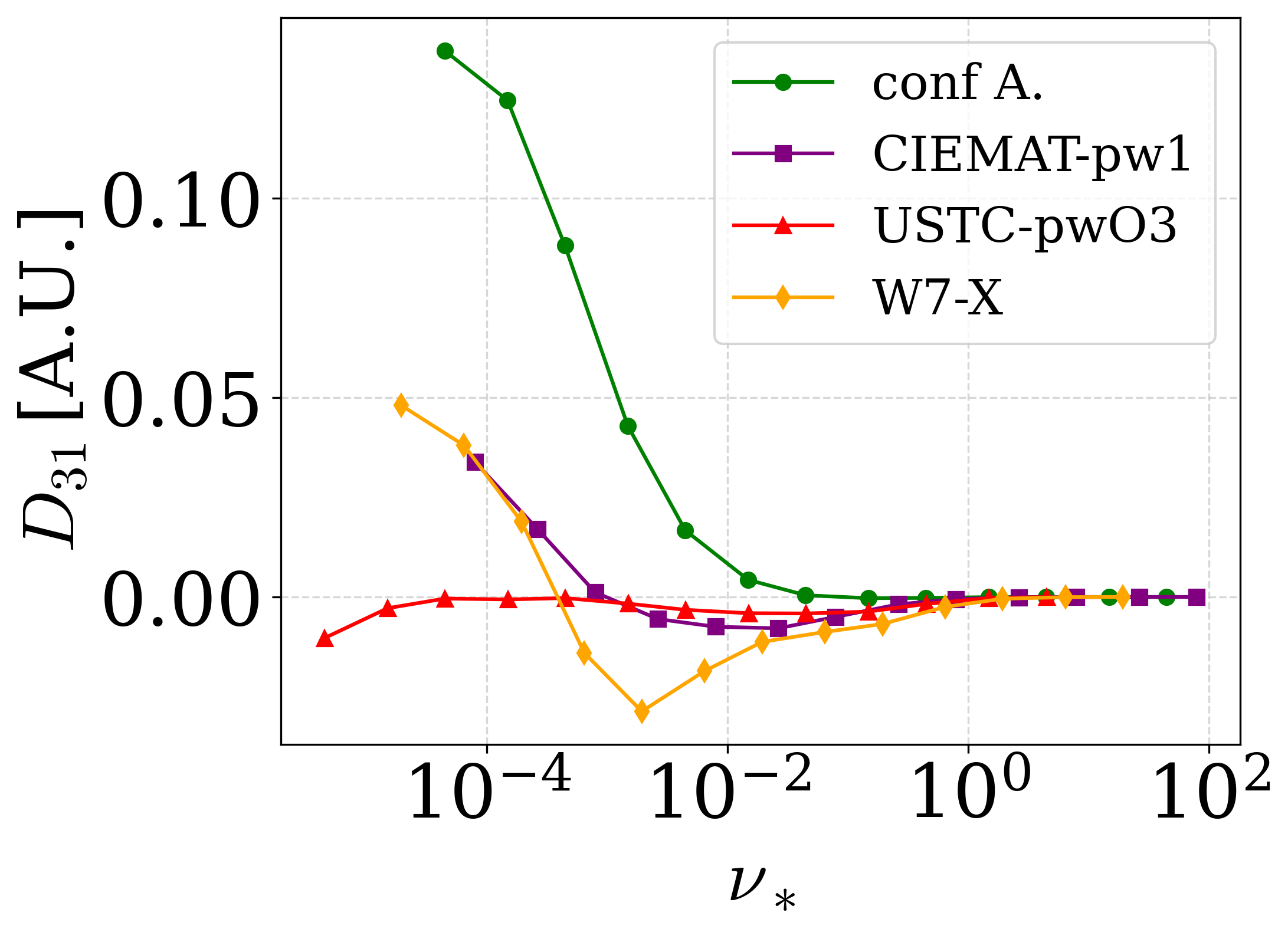}  
\caption{Bootstrap current as a function of the collisionality for the fields of figure~\ref{FIG_CONFS}.\label{FIG_D31CONFS}}
\end{figure}

\section{QI-pwO fields in MHD equilibria}\label{SEC_CONFS}

Once we have validated and employed our model for optimized QI-pwO fields, in this section we present several configurations some of whose flux surfaces can be qualitatively described in this framework. Figure~\ref{FIG_CONFS} shows W7-X high-mirror (top, left), conf. A of~\cite{bindel2023direct} (top, right) and CIEMAT-pw1~\cite{fernandezpacheco2026pw1} (bottom, left), both at $s=0.06$, and USTC-pwO3~\cite{liu2025omni} at $s=0.25$ (bottom, right). At intermediate radial positions, both conf. A and CIEMAT-pw1 are approximately pwO fields very far from quasi-isodynamicity, and the latter has been explicitly optimized to be pwO with a reduced bootstrap current at $s=0.5$~\cite{velasco2024pwO,calvo2025pwO}. However, closer to the axis, where the magnetic field strength can only depend on the toroidal coordinate, both configurations become approximately QI. The pwO configuration of~\cite{liu2025omni}, on the other hand, was optimized to be QI-pwO at $s=1$, although with no requirements on the bootstrap current level. 

While the $B$-maps clearly appear as QI (near $B_\mathrm{min}$) and pwO (near $B_\mathrm{max}$), it is not obvious (recall the scan in $w_2$) that the shape of the pwO piece is appropriate to minimize the bootstrap current. In order to assess this, in each color map, we have overlaid a parallelogram whose shape is described by equations (\ref{EQ_QIPWO}), (\ref{EQ_W1}) and (\ref{EQ_W2}) with $(N_{fp}\zeta_c,\theta_c)=(\pi,\pi)$. The only free parameters, $t_1$ and $t_2$ (which in this section is not determined by equation~(\ref{EQ_T2})), are set in each figure so that two parallel sides of said parallelogram are tangent to \textit{the last} $B$-contour that does not close poloidally (drawn thicker in figure~\ref{FIG_CONFS}). \textit{Indicatively}, one would expect a reduced bootstrap current if the thick $B$-contour fits reasonably well  the parallelogram and, additionally, if the $B\gtrsim B_\mathrm{min}$-contours have the shape required by quasi-isodynamicity.

Figure~\ref{FIG_D31CONFS} indicates that, unsurprisingly, both conditions are relevant. Both conf. A and W7-X present significant deviations from quasi-isodynamicity near $B_\mathrm{min}$; however, the stronger misalignment between the selected $B$-contour and the parallelogram contributes to a larger bootstrap in conf. A. On the other hand, W7-X, CIEMAT-pw1 and USTC-pwO3 display a similar alignment of the parallelogram with the selected $B$-contour; the explicit optimization of the QI part of the flux surface of USTC-pwO3~\cite{liu2025omni} contributes to a smaller value of $D_{31}$ than CIEMAT-pw1 and, specially, than W7-X.


The small bootstrap current of CIEMAT-pw1 and USTC-pwO3 can, to some extent, be traced back to the explicit optimization of their $B$-contours (at outer flux surfaces). On the other hand, the case of W7-X is specially interesting, as its design even predates the demonstration that quasi-isodynamicity can reduce the bootstrap current~\cite{helander2009bootstrap}. The reduction of $D_{31}$ was achieved~\cite{beidler1990w7x,beidler2011ICNTS} by numerically exploring and selecting appropriate amplitudes of the $B_{0,1}$ $B_{1,0}$ and $B_{1,1}$ harmonics in $B(s,\theta,\zeta) = \sum_{m,n} B_{m,n}\cos(m\theta -nN_{fp}\zeta)$. As it turns out, the result was a configuration that is approximately QI, with a deviation that resembles the one proposed in our pwO model for bootstrap reduction.

\section{Discussion}\label{SEC_DISC}

Piecewise omnigenity was presented~\cite{velasco2024pwO} as a concept that could radically expand the range of configurations that can be candidates for fusion reactors. And, indeed, prototypical pwO fields~\cite{velasco2024pwO,calvo2025pwO}, with a spatial variation of $B$ that is qualitatively different to any previous option, have been already shown to be compatible with the plasma physics requirements of a stellarator reactor~\cite{fernandezpacheco2026pw1}. However, as already indicated in~\cite{velasco2025parapwO}, it is likely the case that the best use of this concept is to help improve more mature stellarator strategies towards the stellarator.

Within this perspective, building on piecewise omnigenity and quasi-isodynamicity, we have studied QI-pwO fields. These are pwO stellarator magnetic fields that resemble QI fields in the region $B\gtrsim B_\mathrm{min}$ of the flux surface and present significant deviations from quasi-isodynamicity in the region $B\lesssim B_\mathrm{max}$. By means of a careful tailoring of such deviations, neoclassical radial transport and bootstrap current can be kept at minimal levels. By relaxing the strict quasi-isodynamic constraints, this approach may make it easier to balance neoclassical transport optimization with other engineering and physical considerations in reactor design. Research in this direction is ongoing~\cite{liu2026opwO}.

\begin{acknowledgments}

This work has been carried out within the framework of the EUROfusion Consortium, funded by the European Union via the Euratom Research and Training Programme (Grant Agreement No 101052200 EUROfusion). Views and opinions expressed are however those of the author(s) only and do not necessarily reflect those of the European Union or the European Commission. Neither the European Union nor the European Commission can be held responsible for them. This research was supported by grants PID2021-123175NB-I00 and PID2024-155558OB-I00, Ministerio de Ciencia, Innovaci\'on y Universidades, Spain. The authors are grateful to Eduardo Rogr\'iguez and to the EUROfusion’s TSVV-I team for its valuable feedback, and to all the developers of the code \texttt{MONKES}. Calculations for this work made use of computational resources at Xula (CIEMAT) and Raven (MPG). 

 \end{acknowledgments}

\bibliography{QIpwO.bbl}

\end{document}